%% file: CollegeRankings03-24-2026.tex
\documentclass[12pt]{article}

\usepackage[T1]{fontenc}
\usepackage{authblk}
\usepackage[latin9]{inputenc}
\usepackage[letterpaper]{geometry}
\usepackage{amsmath, amsthm, amssymb}
\usepackage{graphicx}
\usepackage{setspace}
\usepackage{lipsum}
\usepackage{natbib}
\usepackage{comment}
\usepackage{enumerate}
\usepackage{caption}
\usepackage{subcaption}
\newtheorem{theorem}{Theorem} 
\newtheorem{lemma}{Lemma} 
\newtheorem{definition}{Definition}
\newtheorem{corollary}{Corollary}

\newtheorem{proposition}{Proposition}
\newtheorem{claim}{Claim}

\theoremstyle{definition}\newtheorem{example}{Example}
\newtheorem{axiom}{Axiom}

\usepackage{tikz}
\usetikzlibrary{shapes,arrows,automata}
\usetikzlibrary{patterns}

\usepackage{ThayerGraph}
\usepackage{ThayerShortcuts}
\usepgflibrary{shapes.geometric}
\usetikzlibrary{backgrounds}
\usetikzlibrary{snakes}
\usetikzlibrary{arrows,positioning,fit}

\usepackage{pgfplots}
\pgfplotsset{compat=1.17}

\pgfdeclareimage{house1}{house1}
\DeclareMathOperator{\fav}{fav}
\usepackage{bm}
\def\sym#1{\ifmmode^{#1}\else\(^{#1}\)\fi}

\Tball{student}{black}{2}
\Tball{Bigstudent}{green}{4}
\tikzstyle{school}=[rectangle, fill=black!90,inner sep=4pt,draw]
\tikzstyle{student_edge}= [->,color = blue!75, line width = 3pt]
\tikzstyle{red_student_edge}= [->,color = red!75, line width = 3pt]
\tikzstyle{school_edge}= [->,line width = 3pt]
\tikzstyle{invisiball}=[inner sep=0pt,minimum size=0pt,outer sep=0pt]
\Tball{Bigredball}{red}{4}
\Tball{blueball}{blue}{4}
\tikzstyle{black_edge}= [thick,double = black,double distance = 2pt]
\Tball{purpleball}{purple}{2}
\tikzstyle{aaa_black_rect}=[rectangle, fill=black!90,inner sep=4pt,draw]
\tikzstyle{thin_black_edge}= [double = black,double distance = 2pt]
\tikzstyle{dashed_edge}= [thick,double = gray,double distance = 2pt, dashed]
\tikzstyle{student1} = [circle,draw=black,fill=bg_comp2,thick, inner sep=0pt,minimum size=.5cm]
\tikzstyle{student2} = [circle,draw=black,fill=bg_comp1,thick, inner sep=0pt,minimum size=.5cm]
\tikzstyle{school}=[rectangle, fill=beach_green,
		inner sep=7pt,draw, minimum size = 2cm]
\tikzstyle{seat}=[rectangle, fill=beach_green,
		inner sep=3pt,draw, minimum size = .5cm]
\tikzstyle{normal} = [thick, line width = 3pt]
\tikzstyle{point}= [thick, line width=3pt]
\tikzstyle{point2}= [thick,color = bg_comp1, line width=3pt]

\makeatletter
\usepackage{amsfonts}
\usepackage{amsthm}

\usepackage{changepage}
\usepackage{graphics}
\usepackage{color}

\theoremstyle{plain}
\newtheorem*{theorem*}{Theorem}

\makeatother

\usepackage[mathscr]{euscript}
 \let\mathscr\relax
\usepackage[scr]{rsfso}

\newcommand{\tp}{\mathrel{P}}
\newcommand{\tr}{\mathrel{R}}

\newcommand{\bt}{\trianglerighteq}	
\newcommand{\sbt}{\triangleright}
\newcommand{\mus}{\mu^*}

\newcommand{\convp}{\stackrel{p}{\rightarrow}}

\definecolor{darkviolet}{rgb}{0.58, 0.0, 0.83}

\usepackage{hyperref}

    \usepackage[flushleft]{threeparttable}
    \usepackage{booktabs}
    \usepackage{epstopdf}
\usepackage{graphicx}
\usepackage{pdflscape}
\usepackage{rotating}
\usepackage{placeins}

\makeatletter
\newcommand\EightPtClose{\@setfontsize\EightPtClose\@viiipt{9}}
\newcommand\TenPtType{\@setfontsize\TenPtType\@xpt\@xiipt}
\def\notesize{\TenPtType}
\def\notesize{\EightPtClose}
\newenvironment{figurenotes}[1][\vspace{1em}Note]{\begin{minipage}[t]{\linewidth}\notesize{\itshape#1: }}{\end{minipage}}
\makeatother

\begin{document}

\setlength{\parskip}{0pt}
\setlength{\parindent}{15pt}

\begin{titlepage}
\begin{center}

\Huge Desirable Rankings\textsuperscript{*}{\def\thefootnote{}\footnotetext{\textsuperscript{*}Emails: aryalg@bu.edu, tsmorril@ncsu.edu and troyan@virginia.edu. Authors' names appear in random order, as indicated by the $\textcircled{r}$ symbol (AEA confirmation code: fRnaFAunPb1D). We are grateful to Kyle Woodward, who helped us start this project and contributed many thoughtful ideas.  Thayer would like to thank the Stanford market design coffee group, and especially the world's best host, Al Roth, for many helpful suggestions during this project's early stages. We are thankful to Ignacio Willington for his help with the data. We would also like to thank S. Nageeb Ali, Nick Arnosti, Itai Ashlagi, Chris Chambers, Marcelo Fernandez, Ravi Jagadeesan, Daniel Kornbluth, Maciej Kotowski, Andrew Mackenzie, Michael Ostrovsky, Ran Shorrer, Ron Siegel and Alexander Teytelboym, Zitian Wang, as well as audiences at Stanford, Johns Hopkins, UC-Davis, Georgetown, Penn State, Iowa State, Rochester, UConn, UC-Riverside, the United States Naval Academy, the National University of Singapore, the University of Tokyo, London School of Economics, Toulouse School of Economics, Queen Mary University of London, the University of Bristol, University of Michigan, Ohio State, Rutgers University, the Virtual Market Design Seminar, the 2024 Conference on Mechanism and Institution Design, EC22, IIOC 2025, the 12th Conference on Economic Design, and the WUSTL Economic Theory Conference 2026. Troyan gratefully acknowledges support from the Bankard Fund for Political Economy at the University of Virginia.}}

    \Large
    Gaurab Aryal\textsuperscript{1} \textcircled{r}
    Thayer Morrill\textsuperscript{2} \textcircled{r}
    Peter Troyan\textsuperscript{3}
    
    \vspace{0.5em}
    
    \large
    \textsuperscript{1}Department of Economics, Boston University 
    
    \textsuperscript{2}Department of Economics, North Carolina State University 
    
    \textsuperscript{3}Department of Economics, University of Virginia

\vspace{0.25em}

March 22, 2026

\begin{abstract}
We study the aggregation of individual preferences into a collective ranking when agents are matched to alternatives. Examples include college or journal rankings. Our approach rests on a core principle: alternatives that agents \emph{desire} (prefer over their current match) should rank higher socially. We formalize this idea and introduce the Iterated Removal of Underdemanded Schools (IRUS) algorithm to construct these \emph{desirable rankings}. As markets grow large, desirable rankings converge to the true quality ordering of alternatives. We validate this through simulations and an empirical application ranking Chilean medical programs, finding advantages over more traditional revealed preference rankings and Borda counts.
\end{abstract}

\end{center}

\end{titlepage}

\section{Introduction}
People often rely on external rankings of institutional quality to help guide decision-making. Examples include a student choosing a college, a researcher choosing where to submit an article, or a doctor choosing a medical residency program.

One standard approach to constructing a ranking is to use a formula based on indicators of quality. An important example is the \emph{US News and World Report} rankings of United States colleges, which receive a great deal of media attention and have been shown to influence the decisions of college applicants \citep{bowman2009getting, griffith2007influence}. While the precise formula behind this ranking is opaque, the inputs include quality indicators such as the acceptance rate (the percentage of applicants who are accepted) and the yield rate (the percentage of those admitted who enroll). Similarly, researchers often consult journal rankings when choosing where to submit their research. Journal rankings are usually based on acceptance rates or citations. In the National Resident Matching Program (NRMP), which uses a centralized algorithm to match graduating medical students to hospital residency programs, an informal metric often used by programs to judge quality is called the ``rank-to-fill'' ratio: in other words, how far down the program's submitted rank list did they have to go to fill all of their residency slots.\footnote{While this ratio is not an official ranking, many programs care about their rank-to-fill ratios. For instance, \cite{jena2012prevalence} write: ``In our experience, programs that must move farther down their rank list of applicants to fill their openings may be viewed as having an unsuccessful Match; aggressive lobbying of applicants by programs may be a by-product of this motivation.'' \cite{wu2015taking} echo a similar sentiment: ``In a metric-driven world where program quality is gauged by rank-to-fill performance (i.e., ``how low did you go?'') the pressures are manifold.''}

Metrics such as those described above are susceptible to Goodhart's Law:  when a measure becomes a target, it ceases to be a good measure. Indeed, there is evidence that universities purposefully solicit applications from students they know will be rejected to lower their admissions rates, as well as reject highly qualified applicants that they fear may choose another school, so as not to harm their matriculation rates \citep{toor2000pushy,golden2001glass,wsj}.  Similarly, a journal interested in improving its journal impact factor has a strong incentive to manipulate its acceptance rate and citation count, and there are well-known strategies for doing so (see, e.g., \cite{martin2016editors}, \cite{fong2017authorship}, \cite{chapman2019games}, and \cite{ iandt2019}). The ``rank-to-fill'' metric used by medical residencies creates incentives for programs to violate NRMP rules and pressure candidates to reveal their rank-order lists (ROLs) before the match day. The implicit ``threat'' is that if an applicant does not commit to ranking the program highly, the program will demote the applicant in their rankings \citep{sbicca2012follow}, and indeed, many medical students report that pressure from residency programs altered their rankings \citep{jena2012prevalence}.

Rather than relying on seemingly ``objective'' measures such as acceptance rates or citation counts, an alternative approach is to aggregate agent preferences to produce a social ranking. For instance, which colleges students have chosen in previous years can inform the decisions of the current cohort. Information about the best academic journals can be obtained from the order in which researchers submit to (and are rejected from) them. The NRMP collects ROLs of medical students' rankings of residency programs, which can be used to construct a ranking of those programs.

Of course, not all agents will agree on the answers to these questions. Their preferences will be influenced by common factors (e.g., quality, prestige) and idiosyncratic factors (e.g., location, program fit), leading to conflicts in rankings among individual agents. The question then is how to aggregate diverse individual preferences into a single ranking. Many results in social choice theory (e.g., Arrow's Impossibility Theorem \citep{arrow:50}, the Gibbard-Satterthwaite Theorem \citep{gibbard:73,satterthwaite:75}) speak to the difficulty of such preference aggregation.

In this paper, we propose a new method of ranking alternatives. For concreteness, throughout the paper, we refer to the agents as \emph{students} and the alternatives to be ranked as \emph{colleges}. However, our method is general and can be applied to many other settings in which agents have rankings over alternatives. Rather than reducing quality to a formula that assigns ad hoc weights to various characteristics (e.g., admission rates, citation counts), our goal is to use the information contained in students' preference rankings. We seek a method that uses only individuals' ROLs across colleges---a minimal data requirement---and produces an aggregate ranking that reflects the common component of these preferences, filtering out the idiosyncratic component.

In contrast with index measures, our approach does not impose any weights on student or college characteristics, nor does it even need to make choices about what characteristics to include: indeed, student preferences can be influenced by factors that researchers do not easily observe. In contrast to the standard revealed preference-based approach, which is data-intensive and requires carefully chosen utility specifications, the information needed for our ranking algorithm is parsimonious, requiring only the ROLs.\footnote{In an important paper, \label{fn:avery-revealed}\cite{avery2013revealed} propose a \emph{revealed preference ranking} of US colleges using survey data on student preferences, their admission and matriculation choices, and detailed student and college characteristics including tuition. In Section \ref{section:reveal_preference_ranking}, we adapt their framework to our centralized matching data, include additional information on student and college characteristics to capture heterogeneous preferences, and compare the resulting rankings with ours.} Given these ROLs, our ranking algorithm is easy to implement and has a simple, intuitive description that can be easily communicated to non-technical policymakers interested in constructing a ranking.

A feature of our model that we will make use of is that in the markets we consider, there is a matching $\mu$ between students and colleges, where $\mu(i)$ denotes the college that student $i$ attends. Given this, we are motivated by the following intuitive idea: if a student prefers a college $c$ to $\mu(i)$, then this is evidence that $c$ is a better college than $\mu(i)$, and $c$ should also be ranked higher than $\mu(i)$ in the aggregate ranking.\footnote{This is similar in spirit to the revealed preference approach (see footnote \ref{fn:avery-revealed}), which infers that $\mu(i)$ is better than any other college in $i$'s choice set (i.e., any college she was admitted to). The contrast is that we look ``up'' an agent's ROLs to identify schools that are better than $\mu(i)$, whereas revealed preference looks ``down'' to identify schools that are worse than $\mu(i)$.} A second feature that will allow us to attain positive results is that, even though we assume the students have strict preferences, we will not seek a strict aggregate ranking. Indeed, it is probably difficult to discern an exact, strict ranking among colleges that are ``close.'' Instead, we define a ranking as a partition of the colleges into \emph{tiers}, with the interpretation that any two colleges in the same tier are ``tied'' in the rank.\footnote{At the same time, a ranking is only useful if it provides sufficient distinction between colleges. We show both theoretically and empirically that our ranking method produces meaningful separation.}

We begin by providing an axiomatic formalization of our main idea; we call any ranking that satisfies our axioms a \emph{desirable ranking}. Next, we present a simple, intuitive algorithm for constructing desirable rankings. We then prove that any desirable ranking is correct, in the sense that it converges to the true ranking of colleges by quality as the market becomes large. We confirm this result using simulations, which also allow us to explore how the relative weight students place on common vs. idiosyncratic preferences affects the speed of convergence. Finally, we use our method to rank Chilean medical schools. The remainder of the Introduction provides a brief overview of each contribution and concludes with a discussion of additional results.

\textbf{Axioms.} Students may disagree over which college is the best. These disagreements stem from idiosyncratic preferences (e.g., location, fit). Student preferences are also influenced by factors common to all students (e.g., educational quality and prestige). Our goal is to ``filter out'' the idiosyncratic component of preferences and isolate the common component.

The filter we use is Pareto improvement. Consider a matching $\mu$, which is the outcome of some (unmodeled) admissions process. Because colleges have their own selection criteria, there is no reason to expect $\mu$ to be Pareto efficient (from the student's perspective): student $i$ may prefer $\mu(j)$ and vice-versa, which happens when they are each denied admission to the other's college. The key to our approach is to notice that such Pareto inefficiencies reflect underlying differences in \emph{idiosyncratic} preferences. Consider two students, $i$ and $j$, who have the same idiosyncratic preference between two schools, $A$ and $B$. As the common components of their preferences for $A$ and $B$ are also the same (by definition), their overall preferences over $A$ and $B$ will be the same. This means that there can be no Pareto inefficiency, no matter how they are matched. In other words, if $\mu$ is Pareto inefficient, then $i$ and $j$ must have opposing preferences for $A$ and $B$, and one of these preferences is ``reversed'' (relative to the common component) because of idiosyncratic preferences. As the goal of our ranking is to uncover the common component of student preferences, it is more accurate to first Pareto improve $\mu$ to some Pareto efficient $\mu^*$, and then rank relative to $\mu^*$. We call any such Pareto efficient, Pareto improvement of $\mu$ a \emph{shadow matching}. Throughout the paper, we denote shadow matchings with stars. 

Given a shadow matching $\mu^*$, we say that a student \emph{desires} a school $c$ if she prefers $c$ to her shadow matching $\mu^*(i)$.\footnote{
For any initial matching, there may be multiple ways to construct shadow matchings, but, in practice, the choice of shadow matching has a negligible effect on our ranking; see Section \ref{section:MCsimulations}.}  Our first axiom, the \emph{axiom of desire (AoD)}, says that any college $c$ a student desires is ranked higher than $\mu^*(i)$.
This is a natural requirement. Once we have ``filtered out'' idiosyncratic preferences via Pareto improvement, the remaining desire reflects common preferences. Therefore, if a student desires a school $c$ (i.e., she prefers it to her shadow matching $\mu^*(i)$), then this school has a higher common preference component, and so should be ranked higher than $\mu^*(i)$.

Of course, some schools may be low-quality and not be desired by any student. These schools should be ranked low; however, if no student desires a school, then AoD has no bite. What is also needed is a type of converse of AoD that says a school is ranked highly \emph{only if} it is desired by a student at a lower-ranked college. This is the content of our second axiom, which we call \emph{justification}. Another way to state justification is to say that, given some ranking with $K$ tiers, if any lower-tier student does not desire a highly ranked college $c$, then $c$'s initial (high) ranking is not justified, and should be lowered. In particular, any college that no student desires must be ranked in the lowest tier $K$. To justify placing a college $c$ in tier $K-1$, $c$ must be desired by some student in tier $K$, and so on.

\textbf{Ranking algorithm.} We introduce an algorithm for producing rankings that we call \emph{Iterated Removal of Underdemanded Schools (IRUS)}. IRUS is straightforward and can be easily understood pictorially. Consider Figure \ref{fig:desire-graph-intro}, which depicts what we call a desirability graph. Each node represents a school, and we draw a directed arrow from one node to another if a student at the former school desires the latter. We call the schools at the ``bottom'' of the graph (those with no incoming arrows) \emph{underdemanded schools}; in words, a school $c$ is underdemanded if all students weakly prefer their shadow matching to $c$. As we show, for a Pareto efficient matching, the set of underdemanded schools is always nonempty. Intuitively, these schools are the least desirable and should thus be ranked last. So, the algorithm proceeds by identifying these schools and ranking them last. They are then removed, and this process is repeated to get the next-lowest tier, and so on, until all schools have been ranked.\footnote{In the previous version of this paper that considered only unit capacity for all schools, we introduced an algorithm called Delayed Trading Cycles (DTC). DTC is similar in spirit to IRUS: first determine the top trading cycles using \cite{shapley/scarf:74}, then rank them from bottom to top. The DTC ranking also has interesting connections to \cite{richter2015back}, who introduce a concept of `prestige' rankings derived from a definition of competitive equilibrium for abstract economies. Further details on DTC and this connection with competitive equilibrium can be found in our companion paper \cite{aryal2025prices}.} 

Our first main result, Theorem \ref{thm:RUS-is-desirable}, shows that given a shadow matching $\mu^*$, the IRUS ranking satisfies the axiom of desire and justification, i.e., it is a desirable ranking. Further, fixing a shadow matching, the IRUS ranking is the unique desirable ranking relative to $\mu^*$. When the initial matching $\mu$ is itself Pareto efficient, then $\mu$ is its own unique shadow matching, and the IRUS ranking is the unique desirable ranking.

\input{figures/IRUS_desirability_intro}

\textbf{Convergence.} The IRUS algorithm takes a specific shadow matching as an input, and thus, different shadow matchings may lead to different rankings. Our second main contribution shows that these differences vanish as the market becomes large, and all rankings converge to a single ranking; further, this convergence is correct in the sense that it coincides with the true ranking of the colleges based on their underlying quality. Specifically, we consider a model in which a student $i$'s utility for a college $c$ is a convex combination of a vertical component that we call $c$'s \emph{quality}, and that is common across students, and an idiosyncratic component that is specific to $i$ and $c$. Theorem \ref{thm:correct-rankings-limit} shows that as the market grows large, a college with a quality greater than $p\%$ of all colleges is ranked arbitrarily close to the $p^{th}$ percentile. As part of proving this result, we also show that, in the limit, the size of any individual ranking tier becomes small relative to the total market size (Theorem \ref{thm:small-tiers}). Thus, while we do allow for ties in rankings, desirable rankings can still meaningfully distinguish between colleges in different tiers. We also verify this using simulations and in our empirical application. For the Chilean market, with 19-21 medical programs, our ranking produces a meaningful separation of these programs into 12-15 different tiers (depending on the year), with the majority of tiers containing a unique program.

\textbf{Empirical application.} We demonstrate the practical viability of our approach by using it to rank medical schools in Chile. In particular, we use Chilean medical school admissions data from 2006 to 2009 and determine desirable rankings for each year separately. Our desirable ranking method requires data on \emph{only} the observed initial matching and students' ROLs, a particularly parsimonious requirement given that Chilean students submit incomplete lists, ranking on average only three medical programs. In contrast, as we discuss later, the revealed-preference approach requires richer data. Despite this data sparsity and relatively small sample sizes (1,168-1,620 students per year), our IRUS algorithm produces remarkably valid and stable rankings. For instance, Catholic University and the University of Chile are widely considered the top-ranked medical programs in Chile in contemporary national and international rankings. In our desirable ranking, they maintain their first and second positions consistently across all four years. This empirical success validates our approach in a real-world setting, demonstrating that the desirable rankings capture genuine quality differences across medical programs using minimal data. The stability and accuracy of these rankings underscore the robustness of our framework and its practical utility, particularly when traditional quality metrics are unavailable or contested, and when data are limited. It is worth noting that our approach does not require us to take a stand on which student and school characteristics will affect the ROL, nor does it require us to account for out-of-pocket tuition, which we do not observe.  

Although we focus on Chilean medical programs, our approach can be easily applied to other settings with centralized admission systems using ROLs and assignment data. For decentralized markets such as US college admissions, ROLs are less readily available. While this will make calculating our rankings more challenging in practice, the method we introduce is still viable if data on agent preference rankings can be obtained. For example, \cite{avery2013revealed} surveyed students about their preference ordering, application and admission lists, and matriculation decisions. Thus, our ranking methodology is applicable across a variety of different institutional settings.

\textbf{Additional results.} 
To further assess the performance of desirable rankings, we determine two alternative rankings of the medical schools: revealed preference and Borda rankings. We estimate revealed preference rankings by first utilizing \emph{stability} as a solution concept \citep[see][]{gale/shapley:62, Abdulkadiroglu:AER:2003}, which, together with the students' test scores and cutoffs gives us the set of feasible colleges (those with cutoffs that are lower than the student's test score) for each student. Then, we include additional data on students' socioeconomic characteristics and college characteristics, including selectivity measures, but we do not observe tuition. Then, using the discrete choice method, estimate college ``fixed effects,'' which are college qualities that determine the rankings as in \cite{avery2013revealed}. We also determine the Borda rankings, which aggregate students' ROLs by awarding points based on ranking positions \citep{Borda1781}.

These alternative approaches produce rankings that diverge substantially from the well-known rankings of colleges and from our desirable rankings and exhibit notable limitations. The revealed preference method yields volatile year-to-year rankings despite using considerably more data (for instance, Catholic University, which is generally regarded as the top program according to external rankings and is consistently identified as such by desirable rankings, varies across years between 4th and 20th place in the revealed-preference rankings).\footnote{The volatility of revealed preference rankings in our setting is a reflection of data limitations rather than any shortcoming of the approach itself: in the US college context, \cite{avery2013revealed} have access to much more data, most importantly detailed tuition data, that we lack. Our desirable rankings approach, by contrast, succeeds despite using only ROLs and assignment data, combined with a simple algorithm that does not require involved estimation procedures. This makes our approach easier to implement and applicable to many real-world settings where richer data are unavailable.} The Borda rankings, while more stable, still misjudge some key programs and are systematically biased towards programs with larger applicant pools, regardless of quality. These comparisons highlight the better performance of our approach.

\section{Definitions}\label{sec:Model}
There is a set of agents $I=\{i_1,\ldots,i_m\}$ and a set of institutions to which they can be assigned, $C=\{c_1,\ldots,c_n\}$. Throughout the paper, we refer to the agents $I$ as \textbf{students} and the institutions $C$ as \textbf{colleges} or \textbf{schools} for concreteness, though there are many other potential applications. College $c$ has a \textbf{capacity} $q_c\geq 1$.  Each student $i$ has a strict ordinal \textbf{preference relation} $P_i$ over the colleges, where we write $c \tp_i c'$ to denote that $c$ is strictly preferred to $c'$.\footnote{\label{fn:partial-preferences}For most of the paper, we assume that $P_i$ is agent $i$'s true preference ranking. 
There is evidence that individuals may not always submit truthful lists (\citealp{rees2018suboptimal}; \citealp{chen2019self}; \citealp{LarroucauRios2020}; \citealp{hassidim2021limits}; \citealp{shorrer2023dominated}; \citealp{artemov2017strategic,artemov2023stable}), or may not even know their full preferences (\citealp{grenet2022preference}). For example, \cite{Dwenger2018jpube} observe that a nontrivial number of students to German universities submit inconsistent preference lists. Inaccurate or partial preference lists could lead to errors in our rankings. However,  our rankings are robust to some of the most common reasons for partial preferences, such as students submitting short lists because they do not apply to schools they think are unattainable for them, or students who choose to learn their preferences about only a smaller set of `target' schools. Indeed, in the data on Chilean medical programs, students rank an average of only four programs (out of approximately 20), but our rankings still perform well. See Section \ref{sec:empirical} for more details.} We write $R_i$ for the corresponding weak preference relation, i.e., $c \tr_i c'$ if either $c \tp_i c'$ or $c=c'$. Given some subset $C'\subset C$ of the colleges, we use the notation $\fav_i(C')=\max_{P_i} \{c':c'\in C'\}$ to denote $i$'s favorite school in $C'$.

A \textbf{matching} $\mu: I\cup C \rightrightarrows I\cup C$ is a correspondence that satisfies: (i) $\mu(i)\in C$, (ii) $\mu(c)\subseteq I$, (iii) $i \in \mu(c)$ if and only if $\mu(i)=c$, and (iv) $|\mu(c)|\leq q_c$. In words, $\mu(i)$ is the college that student $i$ is matched to, and $\mu(c)$ is the set of students matched to college $c$. Each college $c$ can be matched to no more students than its capacity $q_c$. With slight abuse of notation, given a subset of schools $C'\subseteq C$, we sometimes write $\mu(C')=\{i\in I: \mu(i)\in C'\}$ to denote the set of students who are matched to some school in $C'$. For the purposes of constructing desirable rankings, we treat $\mu$ as exogenously given. Therefore, for most of the paper, we do not need to make any assumptions about how it is determined. For instance, $\mu$ may arise from a decentralized process, as in U.S. college admissions, a more structured process, such as academic journal submissions, or a fully centralized process (as in our empirical application). In Section \ref{section:reveal_preference_ranking}, where we consider revealed preferences ranking as an alternative, we must make additional assumptions on $\mu$, i.e., stability, to identify the underlying preferences and with that the college qualities.

 A matching $\nu$ is a \textbf{Pareto improvement} of matching $\mu$ if $\nu(i) \tr_i \mu(i)$ for all $i$ and $\nu(i) \tp_i \mu(i)$ for some $i$.  If there is no $\nu$ that is a Pareto improvement of $\mu$, then $\mu$ is said to be \textbf{Pareto efficient}. Note that we define Pareto efficiency with respect to student preferences only; this is consistent with our objective to provide a ranking of colleges based on student preferences.\footnote{While we focus on ranking the colleges using the student preferences, given the colleges' rankings of the students, it would also be possible to provide a ranking of the students using the same algorithm we propose.}

A \textbf{ranking} of the colleges is a weak ordering on the set $C$, denoted $\bt$, where $a\bt b$ means that $a$ is ranked (weakly) higher than $b$. If $a\bt b$ but $b \ntrianglerighteq a$,  then we write $a\sbt b$. We allow for ties in rankings, and write $a\simeq b$ when $a\bt b$ and $b\bt a$. 

A ranking $\bt$ can be equivalently written as an ordered partition of the colleges, $$\Pi^{\bt}=\{\Pi^\bt_1, \Pi^\bt_2,\ldots,\Pi^\bt_K\},$$ 
where $\Pi_1^\bt$ is the set of colleges that are ranked the highest, $\Pi^\bt_2$ is the set of colleges that are ranked higher than all others except those in $\Pi^\bt_1$, etc. We refer to the colleges in $\Pi_k^\bt$ as the \textbf{tier-$k$ colleges}, and the students assigned to these colleges, $\mu(\Pi_k^\bt)=\{i : \mu(i)\in \Pi^\bt_k \}$, as the \textbf{tier-$k$ students}. Define the function $\tau:I\cup C\rightarrow \mathbb{N}$ such that $\tau(x)=k$, where $k$ is the tier to which agent $x$ belongs. We will use the letter capital $K$ to denote the lowest tier.

\section{Axioms}\label{sec:axioms}

Our objective is the following: given each student's ordinal ranking of the colleges---which is a combination of common and idiosyncratic preferences---construct an aggregate ranking that reflects the common component. Thus, we would like to ``filter out''  idiosyncratic differences between student preferences before constructing the ranking.

 For illustration, consider the following stylized example: There are $T$ quality levels, where $\gamma(c)\in \{1,\ldots, T\}$ denotes the quality of college $c$. For any two colleges $c',c\in C$, if $\gamma(c')>\gamma(c)$, then $c' \tp_i c$ for all $i$, while if $\gamma(c)>\gamma(c')$, then $c \tp_i c'$ for all $i$. If $\gamma(c')=\gamma(c)$, then, each student $i$ prefers $c \tp_i c'$ with probability 1/2 and prefers $c' \tp_i c$ with probability 1/2. That is, all students prefer higher-quality schools to lower-quality ones, but among schools with the same quality, preferences are drawn randomly. In this example, the quality levels represent the common component of preferences, while the randomness within a quality level represents the idiosyncratic component. As we want our ranking to reflect the common component, it is clear that the correct ranking $\bt$ ranks colleges by quality: $c'\bt c$ if and only if $\gamma(c')\geq \gamma(c)$ (and so also $c' \sbt c$ if and only if $\gamma(c')>\gamma(c)$). Using our tier notation, the correct ranking is such that the ranking tiers are equivalent to the quality levels: $\Pi_t^\bt =\{c\in C: \gamma(c)=t\}$ for $t=1,\ldots, T$. 

Consider a matching $\mu$ in this model. The matching $\mu$ is the outcome of some (unmodeled) admissions process. Because colleges have their own selection criteria in this process, there is no reason to expect $\mu$ to be Pareto-efficient (recall that we define Pareto efficiency from the student's perspective): two students $i$ and $j$ may want to trade. This happens when $\mu(i)$ and $\mu(j)$ are in the same quality level, but $i$ has a higher idiosyncratic (within-quality-level) draw for $\mu(j)$ than $j$ but was denied admission to $\mu(j)$, and vice-versa. In this case, we would have $\mu(j) \tp_i \mu(i)$ and $\mu(i) \tp_j \mu(j)$, and if we were to attempt to construct an aggregate ranking that respected both of these preferences, we run into a contradiction. Notice, though, that this contradiction occurs because $\mu$ is Pareto inefficient, and further, this Pareto inefficiency can be attributed to the idiosyncratic preferences. Indeed, any Pareto improving trades must involve colleges within the same quality level, and not across them. So, rather than ranking with respect to $\mu$, it is more accurate to first Pareto-improve $\mu$ to some Pareto efficient matching $\mu^*$, and then require that if a student prefers $c$ to $\mu^*(i)$, the aggregate ranking should also rank $c \sbt \mu^*(i)$. Notice that this will recover the correct ranking: $c \sbt c'$ if and only if $\gamma(c') > \gamma(c)$. 

Indeed, this insight holds more generally. We will build a formal model later in the paper to make this precise. However, for now, the important takeaway is this: any Pareto inefficiencies that exist at the initial matching $\mu$ reflect underlying differences in idiosyncratic preferences. However, we want our ranking to reflect the common component of student preferences, and not the idiosyncratic component. Pareto-improving upon $\mu$ allows us to filter out the idiosyncratic component of student preferences. We call these Pareto improvements \emph{shadow matchings}. 

\begin{definition}\label{def:shadow-matching}
Fix an initial matching $\mu$. A matching $\mu^*$ is said to be a \textbf{shadow matching} (of $\mu$) if $\mu^*$ is Pareto efficient and $\mu^*(i) \tr_i \mu(i)$ for all $i\in I$. 
\end{definition}
In words, a shadow matching $\mu^*$ is a Pareto efficient, Pareto improvement of the original matching $\mu$.\footnote{Why do we call this a shadow matching? While students cannot trade admissions to a college, if such a (frictionless) ``shadow economy'' were to exist, we would expect the students to make Pareto-improving trades until they were exhausted. The result would be a Pareto efficient, Pareto improvement of $\mu$: the shadow matching $\mu^*$. However, one should not think of these trades as actually being implemented; the shadow matching ``thought experiment'' is simply a tool that we use to produce rankings.} Note that if $\mu$ is Pareto efficient, $\mu$ is also its own shadow matching, and it is the unique shadow matching. When $\mu$ is not Pareto efficient, there may be multiple shadow matchings; however, we will show later that this multiplicity issue is not a significant concern for producing good rankings.

\begin{definition}\label{def:desire}Let $\mu$ be an initial matching and $\mu^*$ be a shadow matching of $\mu$. We say that student $i$ \textbf{desires} a school $c$ if $c \tp_i \mu^*(i)$. Student $i$ \textbf{weakly desires} a school $c$ if $c \tr_i \mu^*(i)$.
\end{definition} 

Desire is a strengthening of simple preference: to desire a school $c$, student $i$ must prefer $c$ not only to $\mu(i)$, but also to the Pareto improvement $\mu^*(i)$. Ranking colleges according to desire is the basis of our approach.

\begin{axiom}[Axiom of Desire]\label{ax:axiom-of-desire}
A ranking $\bt$ satisfies the \textbf{axiom of desire (AoD)} if there exists a shadow matching $\mu^*$ such that for every  $i\in I$ and every $c\in C$ that $i$ desires, we have $c \sbt \mu^*(i)$.
\end{axiom}

We can exhibit the axiom of desire and the role of shadow matchings by returning to our stylized preference example. Consider a case involving four students, $i_1,\ldots,i_4$, and four colleges. Colleges $H_1$ and $H_2$ are high quality, and colleges $L_1$ and $L_2$ are low quality. All students prefer any high-quality college to any low-quality college. Within a quality tier, students have opposing idiosyncratic preferences. This is shown in the preference table in Figure \ref{fig:quality-level-example}.

\input{figures/example_preference_rankings}
 
 The correct ranking in this case is 
 $$
 \{H_1,H_2\} \sbt \{L_1,L_2\}.
 $$ 
 The boxes indicate the matching $\mu$ of students to colleges. Notice that $\mu$ is Pareto-inefficient: for instance, students 1 and 2 would like to trade their assignments because they have opposing idiosyncratic preference draws for high-quality colleges. The matching $\mu^*$ indicated by the stars is a shadow matching of $\mu$. Following the discussion above, our approach is to first Pareto-improve the initial $\mu$ to $\mu^*$ to filter out the idiosyncratic preference draws, and then search for a ranking that satisfies AoD with respect to $\mu^*$. The ranking indicated above satisfies the axiom of desire relative to $\mu^*$: students 3 and 4 both desire $H_1$ and $H_2$, and indeed, these colleges are ranked higher than their respective shadow matchings (students 1 and 2 do not desire anything relative to their shadow matchings, and so the axiom of desire is satisfied trivially).

Our second axiom is a type of converse of AoD. Consider a student $i$ who receives an offer from her top-ranked school $c$. It may be that $i$ is a very good student attending a very good school; alternatively, it may be that $i$ prefers a less good school simply because she has idiosyncratic preferences, and no other students are interested in it (e.g., perhaps $c$ is close to $i$'s home). In the former case, the axiom of desire applies, and $c$ will be ranked highly. However, in the latter case, AoD places no restrictions on where $c$ should be ranked, simply because it is a lower-quality school, and so no student desires it. 

This is highlighted in Figure \ref{tab:jusitified-example}. Students 1, 2, and 3 have the same preferences, and all rank $D$ last. Student 4, on the other hand, has idiosyncratic preferences that rank $D$ first. The indicated matching is Pareto efficient, and thus is the unique shadow matching. Consider the ranking 
$$
\{A,D\} \bt \{B\} \bt \{C\}
$$
This ranking satisfies AoD: notice that school $D$ is ranked (weakly) below the shadow matching of every student, so nobody desires it. This means that AoD has no bite: $D$ can be ranked in any tier without violating it. However, it seems clear here that student 4 most likely has idiosyncratic preferences, and the right ranking for school $D$ is last. 

\input{figures/illustration_justification_axiom}

AoD says that \emph{if} a school is desired by a student, it should be ranked in a higher tier. We also need the converse: a school should be ranked in a higher tier \emph{only if} it is desired by a student from a lower tier. Another way to say this is, if a college is not desired by a student from a lower tier, then its ranking should be lowered; that is, the initial (high) ranking was not \emph{justified}.  The next axiom formalizes this idea. 

\begin{axiom}[Justification]\label{ax:justification}
A ranking $\bt$ is \textbf{justified} if there exists a shadow matching $\mu^*$ such that for every tier $k<K$ and every college $c \in \Pi_k^\bt$, there exists a sequence of colleges $c_k:=c \sbt c_{k+1} \sbt \cdots \sbt c_K$ and a sequence of students $i_{k+1},\ldots,i_K$ such that $c_j\in \Pi^\bt_j$, $\mu^*(i_j)=c_j$ and student $i_j$ desires $c_{j-1}$ for all $j=k+1,\ldots, K$.
\end{axiom}

Justification implies that a college that no student desires must be ranked in the bottom tier $K$. Schools in tier $K-1$ must be desired by some student in tier $K$, schools in tier $K-2$ must be desired by some student in tier $K-1$, and so on. In general, for a school in tier $k$, we must be able to find a chain of students in tiers $k+1,\ldots,K$ such that each student desires the school one tier above their own.

Returning to the Figure \ref{tab:jusitified-example} and the ranking $\{A,D\} \bt \{B\} \bt \{C\}$, school $A$'s ranking is justified, as $A \sbt B \sbt C$, student 2 desires $A$, and student  3 desires $B$. However, school $D$'s ranking is not justified: the only possible sequence is $D \sbt B \sbt C$, but student 2 does not desire $D$. Therefore, $\bt$ is not justified. Indeed, in this example, no student desires school $D$, and so any justified ranking must place school $D$ in the bottom tier.


We call any ranking that satisfies the above two axioms a \emph{desirable ranking}.

\begin{definition}[Desirable rankings]
A ranking $\bt$ is \textbf{desirable} if there exists a shadow matching $\mu^*$ such that $\bt$ satisfies the axiom of desire and is justified with respect to $\mu^*$.
\end{definition}

\section{Computing Desirable Rankings}\label{sec:computing-desirable-rankings}

In this section, we present our main algorithm for computing desirable rankings, called  \emph{Iterated Removal of Underdemanded Schools (IRUS)}. The algorithm is straightforward and works from the bottom up. We begin by identifying a set of schools that are underdemanded, in the sense that no student prefers these schools to their shadow matching. Intuitively, these are the schools that have low desirability and thus should be ranked at the bottom of the ranking. After identifying these schools, we remove them and rank them last. We are then left with a smaller subset of schools and students, which has a new underdemanded set. We identify this set, rank them second-to-last, and remove them. We repeat this process until all schools have been ranked.
\begin{definition}
School $c$ is \textbf{underdemanded} at matching $\mu$ if $\mu(i) \tr_i c$ for all $i \in I$. 
\end{definition}
In words, a school $c$ is underdemanded at a matching $\mu$ if all students weakly prefer their assignment $\mu(i)$ to $c$. 
Though simple, we emphasize the following point, as it is fundamental to our algorithm.  

\begin{lemma}\label{l:ender}
If a matching $\mu$ is Pareto efficient, then the set of underdemanded schools is nonempty.
\end{lemma}

To see this, suppose the set of underdemanded colleges is empty. Select some college $c_1$. Since $c_1$ is not underdemanded, there must exist a student $i_1$ such that  $c_1 \tp_{i_1} \mu(i_1)$.  Set $c_2=\mu(i_1)$ and continue:  $i_2$ is a student who prefers $c_2$ to $\mu(i_2)$, set $c_3=\mu(i_2)$, and so on. This process must cycle, and this cycle is a Pareto improvement.

We are now ready to formally introduce the IRUS algorithm.

\subsubsection*{Iterated Removal of Underdemanded Schools (IRUS)}
Let $\mu^*$ be a shadow matching of $\mu$.
\begin{itemize}
\item Step $\ell =1$: Find the set of underdemanded schools relative to $\mu^*$. Denote this set $C^1$. 
\item Step $\ell$: Remove the schools $C^{\ell-1}$ and the students matched to them, $\mu^*(C^{\ell-1})$. If no students and schools remain, stop. Otherwise, find the set of underdemanded schools relative to $\mu^*$ on the remaining students and schools. Denote this set $C^{\ell}$. 
\end{itemize}
Let $C^1, \ldots, C^L$ be the resulting partition of the colleges. The output of the above algorithm is an \textbf{IRUS ranking} of the colleges, $\bt^{IRUS}$, defined as follows: for any two colleges $a,b \in C$, where $a\in C^\ell$ and $b\in C^{\ell'}$, $a\bt^{IRUS} b$ if and only if $\ell \geq \ell'$. The tier-$k$ colleges are $\Pi^{IRUS}_k=C^{L-k+1}$.

\subsubsection*{Desirability graphs}

The IRUS algorithm is perhaps easiest to understand pictorially, by drawing what we call a \emph{desirability graph}, as in Figure \ref{fig:desire-graph} (desirability graphs were also discussed in Figure \ref{fig:desire-graph-intro} in the Introduction; we reproduce the figure here for ease of reference). We fix a shadow matching $\mus$ and define a graph where each node represents a college and there is a directed edge from college $c$ to $c'$ if a student in $\mu^*(c)$ desires college $c'$.\footnote{Note that we draw an edge if there is at least one student in $\mu^*(c)$ who desires $c'$. This is an important distinction between our approach and alternative ``cardinal'' methods, which might count the number of agents who prefer one school to another, or Borda rankings, which award points depending on how high each school is ranked on an agent's list. Indeed, we believe this feature is one of the reasons to prefer desirable rankings. We discuss this more in Section \ref{sec:alternative-rankings}, where we compare desirable rankings to Borda rankings.} Since the shadow matching is Pareto efficient, this graph is acyclic:  if there were a cycle, implementing the indicated trades would be a Pareto improvement.

\medskip
\input{figures/IRUS_desirability_text}

Suppose that drawing the aforementioned arrows on some market results in the desirability graph depicted in Figure \ref{subfig:initial-desire-graph}. The underdemanded schools are those with no incoming arrows, i.e., those at the ``bottom'' of the graph. This is indicated by the large red stars in Figure \ref{subfig:desire-graph-last-cycles}. IRUS begins by ranking these schools last and removing them. This leaves the desirability subgraph depicted in Figure \ref{subfig:desire-graph-second-to-last-schools}, with a new set of underdemanded schools, again indicated by large red stars. These schools are ranked one step above the schools from the prior step, and are removed. This process is repeated until all schools have been ranked.

\begin{example}\label{ex:running-example}
We provide a small example to show how the algorithm works. There are six students $I=\{i_1,i_2,i_3,i_4,i_5,i_6\}$ and six colleges $C=\{A,B,C,D,E,F\}$. Each college $c$ has a capacity of $q_c=1$. Figure~\ref{fig:irus-step1} shows the student preferences, and the boxes indicate the initial matching $\mu$.

\input{figures/example_IRUS_algo}

\end{example}
The initial matching $\mu$ is denoted by the boxes. We compute the IRUS ranking with respect to the shadow matching $\mu^*$ indicated by the stars in the table. Starting with the entire market in Figure \ref{fig:irus-step1}, notice that $\mu^*(i) \tr_i D$ for all $i\in I$, and so $D$ is underdemaned; the same is true for schools $E$ and $F$. For school $A$, we have $A \tp_1 \mu^*(i_1) = D$, and so $A$ is not underdemanded, nor are schools $B$ and $C$. Thus, the initial set of underdemanded schools is $\{D,E,F\}$, and so we rank these schools last and remove them. 
 
 Figure \ref{fig:irus-step2} shows the submarket after these schools and their students were removed. On this remaining submarket, schools $A$ and $C$ are underdemanded: $\mu^*(i) \tr_i A$ and $\mu^*(i) \tr_i C$ for $i=i_3,i_4,i_5$. But, for school $B$, we have $B \tp_3 \mu^*(i_3)$, and so $B$ is not underdemanded. So, the underdemanded schools are $A$ and $C$, but not $B$. Thus, we rank $\{A,C\}$ one tier ahead of $\{D,E,F\}$. Finally, after removing these schools, the only school left is $B$. The final IRUS ranking is
$$
\bt^{IRUS}: \{B\} \sbt \{A,C\} \sbt \{D,E,F\}.
$$
Notice that this ranking satisfies AoD and justification. For AoD, consider student $i_1$ as an example. Student $i_1$ desires $A$ and $B$, and these are both indeed ranked higher than $\mu^*(i_1)=D$. The same holds for all other students and all schools they desire as well. For justification, student $\mu^*(D)=i_2$ desires both $A$ and $C$, and so it is justified to rank $A$ and $C$ one tier higher than $D$. Similarly, student $\mu^*(C)=i_3$ desires $B$, and so it is justified to rank $B$ higher than $C$. Thus, the IRUS ranking is desirable. 

In general, any IRUS ranking is desirable. We state this formally as the theorem below.

\begin{theorem}\label{thm:RUS-is-desirable}
Any IRUS ranking is a desirable ranking of the colleges. Further, for any shadow matching $\mu^*$, the ranking produced by the IRUS algorithm with input $\mu^*$ is the unique desirable ranking with respect to $\mu^*$. 
\end{theorem}

The proof of this theorem is in the appendix, but the intuition can be seen very clearly using the desirability graph in Figure \ref{fig:desire-graph}. In Figure \ref{subfig:desire-graph-last-cycles}, the large red stars are the bottom tier of the IRUS ranking. When these schools are removed, their students are still pointing ``up'' at the schools they desire. Therefore, schools that they desire have not yet been ranked, and so any desired school is ranked higher, which is the axiom of desire. For justification, the last-ranked schools do not need to be justified, as they are already at the bottom. Consider now the schools at the bottom in the next step (Figure \ref{subfig:desire-graph-second-to-last-schools}). Take any such school $c$. Since $c$ was not removed in the first step, it was not underdemanded. This means that in step 1, there must have been some student $i$ such that $c \tp_i \mu^*(i)$. As $c$ is underdemanded in step 2, $\mu^*(i)$ must have been removed in step 1, i.e., $\mu^*(i)$ is ranked last.   Thus, it is justified to rank $c$ higher than the schools removed in step 1. Repeating this argument shows that the IRUS ranking as a whole is desirable.

Note that when there are multiple possible shadow matchings, there may be different IRUS rankings, depending on which shadow matching is chosen. In the next section, we show that all of these rankings converge as the market grows large. However, there is a special case where there will be a unique desirable ranking, even in small markets: when the initial matching $\mu$ itself is Pareto efficient. 

\begin{corollary}\label{cor:unique-desirable-when-PE}
If the initial matching $\mu$ is Pareto efficient, then there is a unique desirable ranking of the colleges, and it is the ranking produced by the IRUS algorithm.

\end{corollary}

When the initial matching $\mu$ is Pareto efficient, then $\mu$ is its own shadow matching, and further, it is the unique shadow matching. The result above is then an immediate corollary of Theorem \ref{thm:RUS-is-desirable}. This leaves open the question of what to do when there are multiple shadow matchings, and how to ``choose'' the one to use to calculate the IRUS ranking. We turn to this issue in the next section.

\section{Convergence}\label{sec:convergence}

When the initial matching is Pareto efficient, there is a unique desirable ranking. When this is not the case, there could be multiple shadow matchings, which raises the question of how to choose the shadow matching to use when running the IRUS algorithm. While different choices may lead to slightly different final rankings, we show in this section that the precise choice does not have a significant impact, as these differences disappear in large markets. Formally, we show that as the market grows large, all IRUS rankings converge to the same ranking, no matter the choice of shadow matching. Furthermore, we demonstrate that this convergence is to the correct ranking of colleges based on their underlying quality. After showing this result, we provide simulation evidence that allows us to explore the speed of convergence for various parameterizations.

\subsection{Desirable Rankings Are Correct}\label{subsec:rankings-are-correct}
In this section, we show that as the market grows large, any desirable ranking coincides with the true underlying quality rankings of colleges. To make this point formally, we must first expand our model to define what is meant by a college's underlying `quality'. 

Specifically, let each student $i$'s preferences over colleges be determined according to the following utility function:
\begin{equation}\label{eq:student-util-function}
U_{i,c} = \alpha \theta_c + (1-\alpha)\eta_{i,c}.
\end{equation}
$U_{i,c}$ is student $i$'s utility for attending college $c$. The random variable $\theta_c$ represents the intrinsic quality of college $c$. This component of the utility function is the same for all students; thus, $\theta_c$ induces vertical preferences over colleges. The random variable $\eta_{i,c}$ is the idiosyncratic utility specific to student $i$ if she attends college $c$, and thus corresponds to horizontal preferences (e.g., geographic preferences). We assume that all of the random variables are drawn independently from the uniform distribution on $[0,1]$. The uniform distribution is not critical to our results, which continue to hold under general distribution functions (see footnote \ref{fn:observed-heterogeneity}). The number $\alpha\in (0,1)$ represents the weight students assign to college quality relative to the idiosyncratic component. For a student with utility function $U_{i,c}$ we define ordinal preferences in the standard way: $c \tp_i c'$ if $U_{i,c} > U_{i,c'}$ and $c \tr_i c'$ if $U_{i,c} \geq U_{i,c'}$.\footnote{\label{fn:observed-heterogeneity}The IRUS algorithm yields a ranking, and our convergence result provides sufficient conditions under which it converges to the true ranking. In other words, these conditions are model restrictions that help identify the true ranking. We choose the utility specification in (\ref{eq:student-util-function}) and assume $\eta$ is a Uniform random variable for tractability. More generally, we can consider any utility function $U_{i,c}=U(\theta_c,\eta_{i,c})$ and transform the distribution of $\eta_{i,c}$ to Uniform via an appropriate monotonic transformation \citep{lee2016incentive}. 
In practice, there may be observable characteristics of students and colleges that affect preferences and could induce correlation in preferences across students if ignored. For example, students may prefer colleges in their home state, or certain demographic groups may systematically favor particular types of institutions. Our convergence results extend to accommodate such observed heterogeneity by considering $U_{i,c} = \theta_c + X_{ic}^\top \delta + \eta_{i,c}$, where $X_{ic}^\top \delta $ captures utility from observed characteristics $X_{ic}$ (such as college type, state of residence) and $\eta_{i,c}$ represents idiosyncratic preferences. Our convergence result continues to hold so long as the number of types is fixed while the number of students and colleges of each type grows large, and the remaining idiosyncratic preferences $\eta_{i,c}$ are independent across students.}

A correct ranking of colleges should reflect the common quality component, rather than idiosyncratic preferences. That is, we say a ranking is \textbf{correct} if $c \bt c'$ if and only if $\theta_c \geq \theta_{c'}$. Since the random variables $\theta_c$ are uniform[0,1], as the market grows large, a correct ranking will rank a college with quality $\theta_c$ above approximately $(100\times \theta_c) \%$ of the other college. Formally, given a ranking of the colleges $\bt$, we can define the induced \textbf{percentile ranking},
$$\rho(c) := \frac{|\left\{ c'\in C| c \sbt c' \right\}|}{n},$$
where $\rho(c)$ is the percentage of colleges that college $c$ is ranked strictly ahead of. Then, a college with quality $\theta_c$ should have a percentile ranking $\rho(c) \approx \theta_c$.

Of course, it is easy to construct examples of realized random variables where any desirable ranking will not reflect the underlying college qualities. This will occur when idiosyncratic utilities have outsize influence relative to the college qualities (we explore this in our simulations below, where we find that convergence is slower when $\alpha$, the weight on the common component, is smaller). Indeed, this can happen with any ranking system, not just desirable rankings. However, if the market is large, then these events should be relatively rare, and a good ranking method should be able to uncover the underlying college qualities with high probability.

Figure \ref{fig:convergence_intuition} provides intuition of this logic. The left panel shows a generic initial matching $\mu$, in which students are spread across the full support of idiosyncratic preferences. The right panel shows the shadow matching $\mu^*$.
\begin{figure}
\begin{center}
\caption{\bf Schematic Representation of Shadow Matching\label{fig:convergence_intuition}}
\scalebox{0.8}{
\begin{tikzpicture}
\pgfplotsset{
  myplot/.style={
    width=6.5cm, height=6.5cm,
    xmin=0, xmax=1.02,
    ymin=0, ymax=1.02,
    xtick={0,1}, ytick={0,1},
    xticklabels={0,1}, yticklabels={0,1},
    tick style={draw=none},
    axis line style={->, thick},
    xlabel style={font=\large\bfseries},
    ylabel style={font=\large\bfseries},
    tick label style={font=\large},
    axis x line=bottom,
    axis y line=left,
    clip=false,
  },
}
\begin{axis}[
  myplot,
  xlabel={idiosyncratic preference},
  ylabel={college quality},
  at={(0,0)},
  name=leftplot,
    xmin=-0.1, xmax=1.2,
  title={\Large $\mu$},
]
  \draw[dashed, gray, thin] (axis cs:0,1) -- (axis cs:1,1);
  \draw[dashed, gray, thin] (axis cs:1,0) -- (axis cs:1,1);
  \addplot[only marks, mark=*, mark size=4pt,
    mark options={fill=blue!45, draw=blue!60}] coordinates {
    (0.05,0.60)(0.08,0.92)(0.10,0.40)(0.12,0.75)
    (0.15,0.30)(0.18,0.88)(0.20,0.55)(0.22,0.68)
    (0.25,0.20)(0.27,0.45)(0.28,0.82)(0.30,0.35)
    (0.32,0.70)(0.33,0.50)(0.35,0.15)(0.37,0.60)
    (0.38,0.90)(0.40,0.40)(0.42,0.72)(0.43,0.28)
    (0.45,0.55)(0.47,0.85)(0.48,0.18)(0.50,0.65)
    (0.52,0.42)(0.53,0.78)(0.55,0.30)(0.57,0.58)
    (0.58,0.92)(0.60,0.48)(0.62,0.22)(0.63,0.70)
    (0.65,0.38)(0.67,0.62)(0.68,0.82)(0.70,0.25)
    (0.72,0.50)(0.73,0.72)(0.75,0.15)(0.77,0.45)
    (0.78,0.88)(0.80,0.35)(0.82,0.60)(0.83,0.20)
    (0.85,0.75)(0.87,0.42)(0.88,0.95)(0.90,0.30)
    (0.92,0.55)(0.93,0.70)(0.95,0.18)(0.97,0.80)
    (0.98,0.45)(0.10,0.10)(0.50,0.10)(0.70,0.08)
    (0.35,0.95)(0.60,0.95)(0.80,0.98)(0.20,0.98)
  };
  \node[right] at (axis cs:1.22, 0) {\large$\boldsymbol{\eta}$};
\node[left] at (axis cs:0, 1.06) {\large$\boldsymbol{\theta}_c$};
\end{axis}
\hspace{1.0in}
\begin{axis}[
  myplot,
  xlabel={idiosyncratic preference},
  ylabel={college quality},
  at={(7.5cm,0)},
  name=rightplot,
  xmin=-0.1, xmax=1.2,
  title={\Large  $\mu^*$},
]
  \draw[dashed, gray, thick] (axis cs:0,1) -- (axis cs:1,1);
  \draw[dashed, gray, thick] (axis cs:1,0) -- (axis cs:1,1);
  \addplot[only marks, mark=*, mark size=4pt,
    mark options={fill=blue!45, draw=blue!60}] coordinates {
(0.97,0.60)(0.99,0.92)(0.96,0.40)(0.98,0.75)
(0.95,0.30)(0.99,0.88)(0.96,0.55)(0.98,0.68)
(0.94,0.20)(0.97,0.45)(0.99,0.82)(0.95,0.35)
(0.98,0.70)(0.96,0.50)(0.95,0.15)(0.99,0.60)
(0.97,0.90)(0.96,0.40)(0.98,0.72)(0.94,0.28)
(0.97,0.55)(0.99,0.85)(0.95,0.18)(0.98,0.65)
(0.97,0.63)(0.99,0.89)(0.96,0.44)(0.98,0.71)
(0.95,0.33)(0.99,0.85)(0.96,0.58)(0.98,0.64)
(0.94,0.23)(0.97,0.41)(0.99,0.79)(0.95,0.38)
(0.98,0.74)(0.96,0.53)(0.95,0.18)(0.99,0.57)
(0.97,0.87)(0.96,0.43)(0.98,0.69)(0.94,0.31)
(0.97,0.52)(0.99,0.82)(0.95,0.21)(0.98,0.62)
  };
  \node[right] at (axis cs:1.22, 0) {\large$\boldsymbol{\eta}$};
\node[left] at (axis cs:0, 1.06) {\large$\boldsymbol{\theta}_c$};
\end{axis}
\end{tikzpicture}
}
\end{center}
 \begin{figurenotes}
      The left panel shows a schematic representation of an initial matching $\mu$: each point $(\eta_{i,\mu(i)}, \theta_{\mu(i)})$ represents one student $i$'s common and idiosyncratic preferences for her assigned college $\mu(i)$. The right panel shows the same for a shadow matching $\mu^*$. As the market grows large, at the shadow matching, all students are matched to a school for which their idiosyncratic preferences are close to $\eta = 1$ (the maximum possible).
       \end{figurenotes}
\end{figure}
The key observation is that under $\mu^*$, all students are concentrated near $\eta = 1$. To see why this must be so, suppose a student $i$ is matched to college $c$ under $\mu^*$ but has $\eta_{i,c} \ll 1$. Then the student would be willing to trade down to a college with slightly lower quality in exchange for a better idiosyncratic match. As the market grows large, it becomes increasingly likely that there will be another student $j$ at a worse college $c'$ willing to trade for $c$, contradicting Pareto efficiency. Therefore, at $\mu^*$, every student must have $\eta_{i,\mu^*(i)} \approx 1$ for their assigned college. This feature has a sharp implication for desirable rankings: if student $i$ desires college $c$ over college $c'$ under $\mu^*$, it cannot be because $\eta_{i,c} > \eta_{i,c'}$, because both are near $1$. The preference must therefore reflect the quality ordering $\theta_{c} > \theta_{c'}$. Thus, using shadow matching filters the idiosyncratic preferences allowing us to get a quality-based ranking.

We now formalize this intuition. To this end, we study the properties of desirable rankings in a sequence of random markets 
$$
(C^n,\textbf{q}^n,I^{m_n},U^{m_n},\mu^n)_{n=1}^{\infty}
$$
where $C^n=\{c_1,\ldots,c_n\}$ is a set of $n$ colleges, $\textbf{q}^n=(q_1,\ldots,q_n)$ a vector of college capacities, $I^{m_n}=\{i_1,\ldots,i_{m_n}\}$ a set of $m_n$ students, $U^{m_n}$ a vector of student utilities drawn according to equation (\ref{eq:student-util-function}), and $\mu^n$ an initial matching between the students and colleges. For simplicity, we assume that in each market $n$, $\sum_{j=1}^n q_j = m_n$.\footnote{Our results continue to hold whether there is an oversupply and undersupply of college seats.} For each market instance $n$, $\bt^n$ denotes a ranking of the colleges. To simplify notation,  we omit the index $n$ when no confusion arises.

Given $\bt$ and a number $\epsilon > 0$, let
$$
C^\rho(\epsilon) = \{c \in C: \theta_c - \epsilon < \rho(c) < \theta_c + \epsilon\}.
$$
 We think of $\theta_c$ as college $c$'s true or ``target'' ranking, and the set $C^\rho(\epsilon)$ then is the set of colleges whose calculated rankings are within $\epsilon$ of their target. If $c\in C^\rho(\epsilon)$, then we say that college $c$'s ranking is $\epsilon$\textbf{-correct}. Our first main result says that if $\bt$ satisfies the axiom of desire, then for any $\epsilon >0$, as the market grows large, all college rankings are $\epsilon$-correct.

\begin{theorem}[Correct Rankings]\label{thm:correct-rankings-limit}
For each market $n$, let $\bt$ be a ranking that satisfies the axiom of desire, and $\rho(\cdot)$ the induced percentile ranking. Then, for any $\epsilon > 0$, as $n$ grows large, all college rankings are $\epsilon$-correct. Formally:
$$\lim_{n\rightarrow \infty} E\left(\frac{|C^\rho(\epsilon)|}{n} \right) = 1.
$$
\end{theorem}

Notice that the only assumption needed for Theorem \ref{thm:correct-rankings-limit} is the axiom of desire. While justification is not needed for the convergence result, we still view it as an important property for constructing rankings in finite markets. Additionally, Theorem \ref{thm:correct-rankings-limit} applies for \emph{any} ranking that satisfies AoD, i.e., it applies no matter which shadow matching is chosen. Since any IRUS ranking satisfies AoD, we have the following corollary. 

\begin{corollary}
For any $\epsilon > 0$, as $n$ grows large, under any IRUS ranking, the ranking of all colleges is $\epsilon$-correct.
\end{corollary}

\subsubsection*{Overview of the proofs}

The full proof of Theorem \ref{thm:correct-rankings-limit} is in the appendix. Here, we provide an overview of the argument, focusing on the case where $q_c=1$ for all $c\in C$ to exhibit the intuition more easily. 

A key first step in proving Theorem \ref{thm:correct-rankings-limit} is to show that for any desirable ranking, the ranking tiers become small in the limit. We state this result as a theorem because it is also important in its own right, as it allays a potential concern with desirable rankings. The concern is that because we allow for ties, a desirable ranking may not meaningfully distinguish the colleges. Theorem \ref{thm:small-tiers} says that as the market grows large, the size of all ranking tiers, as a percentage of the market size, approaches zero. In other words, desirable rankings do distinguish colleges into different tiers meaningfully. Later, we validate this result both using simulations and in our empirical application, where we show meaningful separation of 19-21 Chilean medical programs into 12-15 different tiers (depending on the year), with the majority of tiers containing a unique program. Formally, we have the following result:

\begin{theorem}[Small tiers]\label{thm:small-tiers}
Let $\bt$ be a ranking that satisfies the axiom of desire. Then, 
$$
\max_k \frac{1}{n}|\Pi^\bt_k| \stackrel{p}{\rightarrow} 0 \text{ as } n\rightarrow\infty.
$$
\end{theorem}

An important corollary of Theorem \ref{thm:small-tiers} that we will use in the proof of Theorem \ref{thm:correct-rankings-limit} is that, if all rankings tiers are small, then the percentage of colleges ranked above (resp. below) any given $\theta\in[0,1]$  converges to $1-\theta$ (resp. $\theta$).\footnote{That $|D(\tilde{\theta})|/n\leq 1-\tilde{\theta}$ follows from the definition of $\rho(\cdot)$. The other direction is not immediate, i.e., it may be that $|D(\tilde{\theta})|/n>1-\tilde{\theta}$. The corollary uses Theorem \ref{thm:small-tiers} to show that if rankings tiers are small, then $|D(\tilde{\theta})|/n$ becomes arbitrarily close to $1-\tilde{\theta}$ as $n$ grows large.}

 \begin{corollary}\label{lem:rankings-convergence}
 Fix $\tilde{\theta} \in [0,1]$, and define $D(\tilde{\theta}) = \{c\in C| \rho(c)\geq \tilde{\theta}\}$ and $D'(\tilde{\theta}) = \{c\in C| \rho(c) < \tilde{\theta}\}$. Then,
 \begin{eqnarray*}
\frac{|D(\tilde{\theta})|}{n}\stackrel{p}{\rightarrow} 1-\tilde{\theta} \quad \text{ and } \quad \frac{|D'(\tilde{\theta})|}{n}\stackrel{p}{\rightarrow} \tilde{\theta}.
 \end{eqnarray*}
 \end{corollary}
 
Theorems \ref{thm:correct-rankings-limit} and \ref{thm:small-tiers} give support for desirable rankings, as they will meaningfully differentiate among the colleges and, with high probability, will uncover the true ranking in the limit.

The proofs of both Theorems \ref{thm:correct-rankings-limit} and \ref{thm:small-tiers} are inspired by an innovative technique of \cite{lee2016incentive}. He uses a result from \cite{dawande2001bipartite} on the size of bicliques in random bipartite graphs to analyze incentives in stable matching mechanisms. We use a similar technique to analyze desirable rankings.\footnote{For another application of Lee's technique to interviews in the NRMP medical residency market, see \cite{echenique2022topofthebatch}.} We first recall some basic concepts from random graph theory.		

A \textbf{graph} is a pair $G= (V, E)$, where $V$ is a set of \textbf{nodes} and $E$ a set of \textbf{edges}. Each edge $e\in E$ is an unordered pair $e=(i,j)$ or $(j,i)$ for $i,j\in V$. A graph $G$ is \textbf{bipartite} if $V$ can be partitioned as $V=V_1\cup V_2 $, where each edge has one node in $V_1$ and one node in $V_2$; for our purposes, $V_1\subseteq I$ will be a subset of the students and $V_2\subseteq C$ a subset of the colleges.

\begin{definition}\label{randomg}
Given a set of nodes $V=V_1 \cup V_2$ and a number $p \in (0,1)$, a \textbf{random bipartite graph} is a graph such that each edge $(i,c)\in V_1 \times V_2$ is included independently with probability $p$. We use $G^p$ to denote a random bipartite graph. 
\end{definition}

A biclique of a random graph is a complete connected subgraph. That is, a \textbf{biclique} is a set of nodes $U_1 \cup U_2$ such that $U_1 \subset V_1$, $U_2 \subset V_2$, and for all $i \in U_1$ and $j\in U_2$, we have $(i,j)\in E$. A biclique has \textbf{size} $a\times b$ if there are $a$ nodes in $U_1$ and $b$ nodes in $U_2$, and is said to be a \textbf{balanced biclique} if $a=b$. We make use of the following result from random graph theory, which says that as $n$ grows large, the size of the maximal balanced biclique in a random graph remains small.

\begin{theorem*}[\cite{dawande2001bipartite}]\label{thm:balanced-biclique-small}
Consider a random bipartite graph $G^p$, where $V=V_1 \cup V_2$ is a partitioned set of nodes, $p\in (0,1)$ is a constant, $|V_1| = |V_2| =n$, and $\beta_n=2 \log(n)/\log(1/p)$. If a maximal balanced biclique of this graph has size $B\times B$, then 
$$
\text{Pr}(\beta_n/2 \leq B \leq \beta_n) \rightarrow 1 \text{ as } n\rightarrow \infty.
$$
\end{theorem*}

We first discuss how we prove Theorem \ref{thm:small-tiers}, as it serves as a stepping stone to proving Theorem \ref{thm:correct-rankings-limit}. The overall strategy is to show that if there is a ``large'' tier, then there is also a ``large'' balanced biclique in an associated random graph. By the theorem of \cite{dawande2001bipartite}, the probability of a large balanced biclique vanishes as $n$ grows large, and thus so also must the probability of a large tier. 

For simplicity, consider the case that $q_c=1$ for all $c$ and $U_{i,c} = \theta_c + \eta_{i,c}$.\footnote{This can be achieved by setting $\alpha = 1/2$ and then rescaling the utility function; in the proof in the Appendix \ref{sec:convergence_proof}, we allow for arbitrary college capacities and arbitrary $\alpha$.} Let $\bt$ be a desirable ranking with shadow matching $\mu^*$ (a Pareto-efficient, Pareto improvement of $\mu$), and consider a tier of colleges $\Pi^\bt_k$. Let $I_k= \{i \in I: \mu^*(i) \in \Pi^\bt_k \}$ be the corresponding students assigned to these colleges at $\mu^*$. We take the colleges in $\Pi^\bt_k$ and divide them into three intervals (Low, Medium, and High) based on their common values (see Figure \ref{fig:random-graph}): 
$$L=\{c \in \Pi^\bt_k: \theta_c<a\} \quad M=\{c \in \Pi^\bt_k: a<\theta_c \leq b \} \quad H=\{c\in \Pi^\bt_k: \theta_c > b\},$$
where $0\leq a < b \leq 1$.  By the axiom of desire, $\mu^*(i)=\fav_i(\Pi^\bt_k)$ for all $i\in I_k$, i.e., each student's most-preferred tier-$k$ college is her shadow matching. We then consider a set of students $U:=\{i\in I_k : \theta_{\mus(i)} \in L\}$.  These students are ``unusual'', in the sense that their most-preferred tier-$k$ college is one with a relatively low common value. Letting $\Delta = b-a$, draw an edge between a student $i$ and a college $c$ if $i$'s idiosyncratic utility for $c$ is low:
$$
\eta_{i,c} < 1-\Delta.
$$
Note that because the $\eta$'s are iid, this draws an edge between each student $i$ and college $c$ independently with probability $p=1-\Delta$, and so this fits into the random bipartite graph model.

\input{figures/random_graph}

Next, notice that $U\cup H$ is a biclique in this random graph. To see why, take some $i\in U$ and $c\in H$, and note that  
$$
\theta_{\mus(i)} + 1 \geq \theta_{\mus(i)} + \eta_{i,\mus(i)} > \theta_c + \eta_{i,c},
$$
where the first inequality follows because $1\geq \eta_{i,\mus(i)}$ and the second follows because $\mus(i)$ is $i$'s most-preferred tier-$k$ college. The first and last inequalities can be rearranged to 
$$
1-(\theta_c - \theta_{\mus(i)}) > \eta_{i,\mus(i)}.
$$
As $\theta_c> b$ (because $c \in H$) and $\theta_{\mus(i)}\leq a$ (because $\mus(i)\in L$), we have  $\Delta < \theta_c - \theta_{\mus(i)}$, and so
$$
1-\Delta > \eta_{i,\mus(i)},
$$
which is precisely our condition for an edge above. Thus, there is an edge between every student in $U$ and every college in $H$, i.e., $U\cup H$ form a biclique. This means there is a balanced biclique of size at least $B=\min\{|U|,|H|\}$.\footnote{Technically, the balanced biclicque has size $B\times B$; for simplicity, we slightly abuse notation and refer to $B$ as the size of the biclique.} Further, we have $|U|=|L|$ (by the definition of $U$), and so there is a balanced biclique of size $B=\min\{|L|,|H|\}$. The remainder of the proof shows that because the college qualities are drawn iid uniform$(0,1)$, as the market grows large, with high probability, we can always find some interval $[a,b]$ such that at least 1/4 of the colleges have common value less than $a$, and at least 1/4 of the colleges have common values above $b$. This implies that $B$ is large (on the order of $n$), in violation of \cite{dawande2001bipartite}, which says that the size of the maximum balanced biclique grows only on the order of $\log(n)$.

After showing Theorem \ref{thm:small-tiers} and its corollary, Corollary \ref{lem:rankings-convergence}, we use them together with the random graph result again to prove Theorem \ref{thm:correct-rankings-limit} as follows. For any $\theta \in [0,1]$ and $\epsilon>0$, define two sets of colleges:
$$
L=\{c\in C| \theta_c\geq \theta \text{ and } \rho(c)\leq \theta-\epsilon\}\quad \text{ and } \quad W=\{c\in C| \theta_c<\theta' \text{ and } \rho(c) > \theta - \epsilon\},
$$
where $\theta-\epsilon < \theta' < \theta$. In words, $L$ is a set of colleges that ``lose'' in the ranking in the sense that they are at least $\epsilon$ below their target, and $W$ is a set of colleges that are ``wrongly ranked'' in the sense that their quality is (relatively) low, but they are ranked above any college in $L$. 

The goal is to show that $|L|/n \stackrel{p}{\rightarrow} 0$. We once again define a set of students,
$$
U=\{i\in I: \mus(i)\in W\}
$$
who are ``unusual'', in the sense that their shadow matchings are relatively overranked. We construct another bipartite graph, connecting a student $i$ and a college $c$ if
$$
\eta_{i,c}<1-(\theta-\theta')
$$
By AoD, for any $i \in U$ and $c\in L$, we have $\mus(i) \tp_i c$, which can be used to show that $\eta_{i,c}<1- (\theta-\theta')$. This is precisely the condition for an edge in our graph, and so the set $L\cup U$ forms a biclique. Because $|U|=|W|$, there is a balanced biclique of size 
\begin{eqnarray}\label{eq:min-L-W-to-0}
\min\left\{\frac{|L|}{n},\frac{|W|}{n}\right\} \stackrel{p}{\rightarrow} 0,
\end{eqnarray}
where the ``$\convp$'' follows from \cite{dawande2001bipartite}. The last step is to show that $|W|/n\convp \omega$ for some $\omega > 0$. Define $A=\{c\in C| \rho(c)>\theta-\epsilon\}$. By Corollary \ref{lem:rankings-convergence} to Theorem \ref{thm:small-tiers}, $|A|/n \convp 1-(\theta-\epsilon)$. Finally, because $\theta-\epsilon < \theta'$, a non-vanishing proportion of the colleges in $A$ must have qualities $\theta_c$ less than $\theta'$, which is the set $W$.\footnote{Because the $\theta_c$'s are uniformly distributed, as $n$ grows large, the proportion of colleges with $\theta_c \geq \theta'$ approaches $1-\theta'$. Because $1-\theta' < 1-(\theta-\epsilon)$ and $|A|/n \convp 1-(\theta-\epsilon)$, intuitively, in order to ``fill'' the set $A$, it must include some nontrivial proportion of colleges with qualities less than $\theta'$.} Thus, $|W|/n \convp \omega >0$, which, combined with (\ref{eq:min-L-W-to-0}) implies that $|L|/n \convp 0$

The above is an outline of how we show that the proportion of colleges with ranking $\rho(c)$ more than $\epsilon$ below their target becomes vanishingly small as $n$ increases. We also must show that the proportion of colleges whose ranking $\rho(c)$ is more than $\epsilon$ above their target becomes vanishingly small as well. The argument proceeds similarly, and the details can be found in the full proof of Theorem \ref{thm:correct-rankings-limit} in the appendix.

\subsection{Simulations \label{section:MCsimulations}}
Having established that the desirable ranking converges to the true quality ordering, we now use simulations to evaluate the performance and convergence properties of desirable rankings. Following the theoretical framework, we begin with the utility function in Equation (\ref{eq:student-util-function}), and systematically vary the market size $n\in\{10, 50, 100, 200, 500\}$ and the relative importance of objective quality versus individual taste through the parameter $\alpha\in\{0.01, 0.2, 0.6, 0.9\}$.

We set each market to have $n/2$ colleges, each with two seats, and ensure that every student receives an assignment. Smaller values of $\alpha$ indicate that students rely more heavily on idiosyncratic preferences, making the recovery of the true rankings more challenging and representing a more demanding test of our algorithm.

For each combination of $n$ and $\alpha$, our simulation procedure follows four steps: (i) generate college qualities $\theta_c$ and student idiosyncratic preferences $\eta_{ic}$ as independent uniform random variables; (ii) use the random assignment to determine the initial matching $\mu$; (iii) create 100 shadow matchings $\mu^*$ and implement our IRUS algorithm for each; and (iv) compute percentile rankings $\rho$ for each parameter combination and shadow matching.\footnote{To generate $\mu^*$ in step (iii), we apply a ``multi-capacity version'' of the Top Trading Cycles (TTC) mechanism in \cite{shapley/scarf:74}. We first randomly order the students matched to each college under $\mu$. Then colleges point to their highest-ordered remaining student, with seats decreasing by one and students removed as cycles form. To generate 100 shadow matchings $\mu^*$, we change the random ordering 100 times and compute the resulting TTC cycles.\label{footnote:ttc}} We simulate these steps 1,000 times to generate our sample.

\input{figures/scatter_average_bias}

Figure \ref{fig:scatter_rho_true} visualizes the performance of desirable rankings across different parameter specifications. We present results for one randomly selected (93rd) shadow matching, with the remaining 99 cases shown as a grey band. The scatter plots show that as market size ($n$) increases and the preference weight on true quality ($\alpha$) grows, the mean percentile rankings converge to the true college rankings, with data points clustering around the 45-degree line. Desirable rankings demonstrate good performance, achieving a near-perfect ranking in larger markets with higher $\alpha$. The narrow (almost imperceptible) grey bands across all parameter combinations confirm the robustness of our approach to the choice of shadow matching $\mu^*$.

Table \ref{tab:bias} reports three standard performance measures. \textit{Bias} measures the average difference between estimated percentile rankings $\rho$ and true rankings $\theta$. \textit{Variance} captures the stability of our rankings. \textit{Mean squared error (MSE)} combines bias and variance into an overall performance measure.
We compute these metrics across 1,000 simulations for the same 93rd shadow matching shown in Figure \ref{fig:scatter_rho_true}. 
The desirable ranking improves with both market size and $\alpha$. Even with $\alpha = 0.01$, MSE decreases from 0.313 to 0.088 as market size grows from 10 to 500 students. When students strongly value quality ($\alpha = 0.90$), performance is excellent with MSE dropping to just 0.001 with 200 students.

\input{tables/simulation_bias}

The ranges [minimum, maximum] below each entry show statistics across all 100 shadow matchings, which are remarkably tight. For example, with $\alpha = 0.20$ and $n = 100$, bias ranges only from 0.050 to 0.052, demonstrating that different shadow matchings yield essentially identical rankings.

Figure \ref{fig:uniform_convergence} examines uniform convergence using the supremum norm $\|\rho - \theta\|_{\infty}:=\sup_{c\in C}\|\rho(c) - \theta_c\|$, defined as the supremum absolute difference between percentile ranking and true rankings across all colleges within each simulation. For each $(n,\alpha)$, we calculate the supremum norm for each of the 1,000 simulations, and then report the mean and median across simulations.

The figure shows clear convergence patterns, with both mean and median supremum norms decreasing systematically with market size. As expected, convergence is faster for higher values of $\alpha$. The gray bands represent the minimum and maximum mean supremum norms across all 100 shadow matchings and are barely visible, indicating consistent performance regardless of which shadow matching is used.

The simulations demonstrate that desirable rankings provide a reliable method for inferring college quality from ROL and $\mu$ because they are consistent (performance improves systematically with market size), informative (performance improves with weights on true quality), and robust (rankings remain stable across shadow matchings). 

\input{figures/simulation_uniform_convergence}

\section{Application: Ranking Chilean Medical Colleges\label{sec:empirical}}

Chile's centralized college admission system provides an ideal setting to apply our ranking methodology. The system is transparent, and we have access to comprehensive data, including students' ROLs, enabling the implementation of the IRUS algorithm.

The most selective universities in Chile participate in the centralized admission system managed by DEMRE (Departamento de Evaluaci\'on, Medici\'on y Registro Educacional). DEMRE implements a college-proposing deferred acceptance algorithm to match high school graduates with college programs. The process begins each December when students take the PSU (Prueba de Selecci\'on Universitaria), a standardized test with mandatory components in Language and Mathematics, plus optional tests in Sciences and Social Sciences. PSU scores range from 150 to 850, standardized annually to approximate a Gaussian distribution with a mean of 500 and a standard deviation of 110.

Students' academic profiles incorporate their high school GPA (PEM - Notas de Ensea\~nza Media) and within-school ranking. Each program sets admission criteria by specifying weights for these components and establishing minimum PSU thresholds. In early January, students submit a single application listing up to eight programs in strict preference order. For each student-program pair, a weighted admission score is calculated by combining PSU results, GPA, and school ranking using program-specific formulas. The deferred acceptance algorithm then proceeds iteratively until all eligible students are matched or exhaust their preference lists, with each student receiving at most one offer.

\subsection{Data}

We use administrative data from Chile's centralized admission system for the years 2006-2009, focusing exclusively on medical and dental programs, which we refer to as medical programs. This restriction ensures our analysis examines a coherent set of highly competitive professional programs with similar application patterns and career outcomes.

Our sample contains a total of 5,411 students who applied only to the 22 medical programs operating during this period. Medical programs differ in terms of the degrees, the name of the university, and their locations. The annual student counts are 1,302 (2006), 1,321 (2007), 1,168 (2008), and 1,620 (2009). The data include information on student demographics, test scores, high school characteristics, program-specific weighted scores, ROLs, and admission outcomes.

Table \ref{tab:summary_statistics} presents summary statistics for student and college characteristics. Panel A shows that medical program applicants represent a highly qualified and homogeneous population. Average PSU scores exceed 6,700 points.\footnote{Individual test scores (e.g., language and math PSU tests) range between 150 and 850. However, when DEMRE (\emph{Departamento de Evaluaci\'on, Medici\'on y Registro Educacional}) reports average PSU score across tests, it scales it up by a factor of ten to avoid decimals, reporting in thousands. When it reports the weighted averages that each program computes for each applicant--combining test scores with high school GPA according to program-specific weights--as well as the program-level cutoffs are in ten-thousands. Throughout the analysis, we follow DEMRE's original coding and retain all score variables in the units as reported in the data.} Nearly all applicants are unmarried, with approximately 40\% from the Santiago metropolitan region. Among the students, over 40\% have university-educated parents, nearly 60\% have private health insurance, and fewer than 20\% attend municipal (public) schools.

 \input{tables/data_summary_stat} 
     
Panel B presents college-level characteristics across all the medical programs. Admission cutoff scores cluster tightly around 71,500 points, reflecting the uniformly high standards of medical education in Chile. The geographic distribution shows that most programs are located in the southern regions (62\%), with central Chile accounting for 29\% and the northern regions containing the remainder.

We also construct and present the average and standard deviation of three complementary (standardized) selectivity measures to capture different dimensions of institutional competitiveness. Cutoff-based selectivity is constructed by collecting the median of the last cutoff scores (minimum admission scores) for each program across all ranking positions where it appears in student ROL, reflecting the academic threshold required for admission. Preference-based selectivity is calculated by finding the average ranking position where each college appears in students' ROLs (across positions 1-8), then transforming this as (maximum rank + 1 - average rank) so that colleges typically ranked first receive higher selectivity scores than those ranked lower, capturing revealed preferences for institutional desirability independent of admission requirements. Student quality-based selectivity measures the mean PSU test scores of students who enrolled at each college, reflecting the caliber of the student body. 

All three selectivity measures are standardized to have a mean of zero and a unit variance, making them comparable both within and across years despite different original scales. Higher values indicate a more selective program across the admission threshold, student preference, and aptitude of the enrolled students. The fact that the average selectivity measures are close to zero suggests that all programs are homogeneous, making ranking a challenging task.    

As we can see from Panel A, students submit relatively short preference lists, ranking an average of 3.5 to 3.9 programs despite being allowed up to eight choices. This reflects both the specialized nature of medical education and students' tendency to focus their applications on programs where they have realistic admission prospects. The variable ``Which Ranked College'' indicates the position on students' ROLs where their enrolled program appeared, with lower numbers indicating higher-ranked preferences. Students, on average,  match with one of their top two programs.  

This pattern bears out in Table \ref{tab:match_outcomes}, which presents admission outcomes by ranking position. Admission rates range from 37.28\% (2009) to 44.16\% (2006), with most admitted students receiving their first or second choice. The concentration of matches at top-ranked positions reflects both the quality of student applications and the effectiveness of the preference revelation process. Students who receive offers lower on their lists typically applied to more competitive programs initially.

\input{tables/data_medical_school_matches}

Our analysis focuses exclusively on students who were both admitted and enrolled, excluding those who received offers but chose not to enroll in any program. This restriction ensures the cleanest possible test of our ranking methodology.

\subsection{Desirable Rankings}

To construct the desirable ranking of Chilean medical programs, we use only the ROLs and admission data and apply the IRUS algorithm. As mentioned above, we focus exclusively on students who ranked only medical programs and were both admitted and enrolled in one medical program.

The implementation follows a multi-step process, separately for each year. We begin with the ROL and admission data, which we denote as the initial assignment $\mu$, and add information about program capacities calculated from the enrollment numbers. We then apply a multi-capacity version of the Top Trading Cycles mechanism to generate shadow matching $\mu^*$ (see Footnote \ref{footnote:ttc}). Finally, we implement the IRUS algorithm to identify sequences of underdemanded programs and construct the desirable rankings.

\input{tables/desirable_rankings}

Table \ref{tab:desirable_rankings} shows the desirable ranking of all medical programs by year. The first thing to note is that the IRUS ranking delivers significant separation of programs into distinct tiers. With 19-21 programs (depending on the year), the rankings separate them into 12-15 different tier levels. The modal tier has one program, with a maximum tier size of 3, and in any given year, there are at most two tiers that have more than two programs (in 2006, all tiers have a size of either 1 or 2). This finding is in line with Theorem \ref{thm:small-tiers}---which, recall, says that the size of the largest tier remains small as the market grows large---even in a market of modest size.

Furthermore, the top-ranked programs demonstrate remarkable stability despite our relatively small sample sizes and parsimonious specification. The medical program offered by Catholic University (ID 1258) consistently ranks first across all four years, followed by the University of Chile's medical program (ID 1183), which maintains second place across all four years. The University of Concepci\'on (ID 1386) and the University of Santiago de Chile (ID 1691) also show consistent top-rank performance. This stability is particularly noteworthy, given that our rankings are derived purely from observed ROL, without requiring students to explicitly state their preferences for program characteristics. The consistency suggests that underlying program quality differences are persistent and well-recognized by students.

Our theoretical framework provides additional confidence in these results. Theorem \ref{thm:correct-rankings-limit} establishes sufficient conditions under which the IRUS algorithm recovers the true underlying quality ranking, ensuring that our empirical findings reflect genuine program quality differences. 
Although our data span 2006-2009, and we do not have access to the prevailing rankings from those years, current widely used rankings show remarkable alignment with our results from 2006-2009. For instance, the QS World University Rankings and Universo Educativo, a Chilean educational platform, consistently rank the Catholic University of Chile and the University of Chile as the top medical institutions in the country.\footnote{See, ``QS World University Rankings by Subject: Medicine,'' 2025, \url{https://www.topuniversities.com/university-subject-rankings/medicine}. When restricted to Chile, the programs in our sample are ranked in the following manner: Catholic University of Chile, University of Chile, University of Santiago of Chile, and Austral University of Chile.}$^,$\footnote{Universo Educativo, ``Top 10 Universidades de Medicina en Chile | Ranking 2025,'' July 10, 2024, \url{https://universoeducativo.cl/universidades/medicina/}, which ranks the programs in our sample in the following order: Catholic University of Chile, University of Chile, University of Concepc\'ion, University of Santiago de Chile, Austral University of Chile, University of Valpara\'iso, University of La Frontera.} 
Similarly, national entrance examination data reveal that the Catholic University of Chile attracted the majority of the top 100 students (based on PAES scores) in 2025, demonstrating continued alignment of student preferences with our ``historical'' desirable rankings.\footnote{PAES (Prueba de Acceso a la Educaci\'on Superior) is the current standardized college entrance examination that replaced the PSU (Prueba de Selecci\'on Universitaria) in 2022. See Pontificia Universidad Cat\'olica de Chile, ``L\'ideres en selecci\'on PAES 2025: 51\% de los 100 mejores puntajes estudiar\'a en la UC,'' January 20, 2025, \url{https://www.uc.cl/noticias/lideres-en-seleccion-paes-2025-51-de-los-100-mejores-puntajes-estudiara-en-la-uc/}. } These findings demonstrate the applicability and empirical content of our methodology.

The accuracy extends meaningfully beyond the top-tier programs. The University of Concepci\'on has maintained its top-three position, as predicted by our rankings, while the University of Santiago Chile currently ranks fourth, precisely within its  3-4 range from our desirable rankings. Regional institutions, such as the Austral University of Chile and the University of La Frontera, continue to occupy the positions our methodology identified as ``lower'' relative to Santiago-based elite institutions. This middle-tier predictive success demonstrates that the IRUS algorithm captures not only absolute institutional quality but also relative competitive positioning within the Chilean medical education landscape, achieving reasonable accuracy in predicting contemporary hierarchical positions among established institutions.\footnote{While our desirable rankings are accurate for established institutions, one of the main reasons the recent ranking differs from our ranking is the emergence of high-performing private universities that entered the market after our study period. For example, the University of Los Andes and the University of Desarrollo are now considered among the top (sixth and seventh) medical schools, but were absent from our sample. See, the rankings by Universo Educativo here \url{https://universoeducativo.cl/universidades/medicina/}.}

\section{Alternative Rankings}\label{sec:alternative-rankings}
To evaluate our rankings against other alternatives, we construct two additional rankings using the Chilean data and compare their performance against our approach. We begin with the revealed preference rankings and then turn to the Borda rankings.

\subsection{Revealed Preference Rankings\label{section:reveal_preference_ranking}}

To construct revealed-preference rankings, we estimate college quality from our admissions data and rank programs based on these quality measures. The revealed preference approach employs a discrete choice framework to estimate college qualities in the form of college fixed effects \citep{avery2013revealed}. Unlike our approach, this methodology requires additional data on student socioeconomic background, their scores, college attributes, program cutoffs, and tuition costs, and imposes additional structure on the utility function, including specifying which student and college characteristics enter preferences and which do not. \cite{avery2013revealed} can credibly implement this approach, as they have access to richer data: complete and truthful ROLs for all students, the set of admitted colleges, and detailed tuition information. We do not have comparable data in the Chilean setting, which makes direct application of their approach infeasible.

A natural alternative would be to use students' observed ROLs directly for estimation. However, in the Chilean system, students rank up to eight programs and typically do not exhaust all available slots. This restriction means that observed ROLs are incomplete and likely reflect strategic considerations, such as omitting programs perceived as out of reach. As \cite{fack2019collegechoice} show, relying on incomplete and strategically selected ROLs in an exploding logit model yields inconsistent estimates.

Instead, we use \emph{stability} as our identification strategy. Since the Chilean system employs a college-proposing Deferred Acceptance mechanism, the observed matching must satisfy stability \citep{gale/shapley:62, Abdulkadiroglu:AER:2003}. Stability requires that no student-college pair exists such that both the student prefers the college to their current assignment and the college prefers the student to at least one of its current admits. In other words, every student is matched with their favorite feasible college, where a college is feasible for a student if its ex post cutoff score is below the student's score. These cutoffs and scores are recorded in our data, allowing us to determine a set of feasible programs for each student. Then the stability condition effectively implies a discrete choice framework in which each student faces a personalized choice set of feasible colleges.

{\bf Utility Specification.} Let student $i$'s utility from attending college $c$ be given by
\begin{equation}
U_{i,c} = \theta_c + \delta_{i,c} + \eta_{i,c} \equiv \theta_c + \delta(X_i, W_{i,c}; \boldsymbol{\beta}) + \eta_{i,c}, \label{eq:utility}
\end{equation}
where $\theta_c\in\mathbb{R}$ is the intrinsic college quality, $\eta_{i,c}$ is student $i$'s idiosyncratic preference shock, and $\delta(\cdot)$ is a deterministic utility component that depends linearly on student characteristics $X_i$, student-college match characteristics $W_{i,c}$, and is known up to a vector of parameters $\boldsymbol{\beta}$.

We assume that preference shocks are independently and identically distributed according to a Gumbel distribution: $\eta_{i,c} \sim^{i.i.d.} \text{Gumbel}(0,1)$. Let $\boldsymbol{W}_i \equiv \{W_{i,c}\}_{c\in \mathcal{C}}$ and $\boldsymbol{\eta}_i \equiv \{\varpi_{i,c}\}_{c\in \mathcal{C}}$ denote the collection of all match characteristics and preference shocks for student $i$, where $\mathcal{C}$ is the set of all available colleges. We assume independence between preference shocks and observables: $\boldsymbol{\eta}_i \perp (X_i, \boldsymbol{W}_i)$.

We further assume that all students must enroll in some program, so there is no outside option. 
Furthermore, we assume that, conditional on student and college characteristics, there are no peer effects: a student's preferences over colleges are not affected by other students' assignments or by equilibrium outcomes such as admission cutoffs.

{\bf Stability Criteria.}
We now assume that the matching $\mu$ is stable. Let $P(\mu)$ be the cutoffs associated with $\mu$, which are random variables determined by the unobserved utility shocks $(\boldsymbol{\eta})$. Let ${\boldsymbol s}_i$ be student $i$'s score. Matching $\mu$ is the outcome of a discrete choice model with a personalized feasibility set, $\mathcal{C}({\boldsymbol s}_i, P(\mu))$. The conditional probability that $i$ is matched with college $c$, or chooses $c$ in $\mathcal{C}({\boldsymbol s}_i, P(\mu))$, is
\begin{align*}
\Pr\!\left(c = \mu(u_i, {\boldsymbol s}_i) = \arg\max_{c' \in \mathcal{C}({\boldsymbol s}_i, P(\mu))} U_{i,c'} \mid (X_i, W_{ic}), {\boldsymbol s}_i, \mathcal{C}({\boldsymbol s}_i, P(\mu)); (\boldsymbol{\theta}, \boldsymbol{\beta})\right).
\end{align*}

We further assume that, conditional on observables, student preferences and scores are independent of the set of feasible colleges. This assumption is reasonable given the size of the market, as individual students cannot affect cutoffs by changing their scores. Given these assumptions, the corresponding conditional log-likelihood function is
\begin{eqnarray*}
\hspace{-0.5in}\ln L(\boldsymbol{\theta, \beta} \mid \mathbf{(X,W)}, \mathbf{s}, \mathcal{C}({\boldsymbol s}_i, P(\mu))) = \sum_{i=1}^I \sum_{c=1}^C \delta_{i,c} \times \mathbf{1}(\mu(u_i, {\boldsymbol s}_i) = c) - \sum_{i=1}^I \ln\!\left(\sum_{c' \in \mathcal{C}({\boldsymbol s}_i, P(\mu))}\!\!\!\!\!\!\!\!\theta_{c'}+ \exp(\delta_{i,c'})\right).
\end{eqnarray*}

These feasible sets differ across students, which helps identify the college quality. This identification strategy requires two key assumptions: first, that a student's priority ranking does not directly influence their utility from attending a college, and second, that students do not experience spillover effects from their peers' choices. For a detailed discussion of these identifying conditions, which builds on \cite{Matzkin1993}, see \cite{fack2019collegechoice}.

To control for student heterogeneity, we include a set of demographic and socioeconomic characteristics in the utility function. These characteristics include gender, secondary school type (private or municipal), secondary school location, work status, family income bracket, father's education level, type of health insurance, and standardized test scores. For program characteristics, we use all variables in Summary Table~\ref{tab:summary_statistics}, including three selectivity measures that we constructed: separate indicators for first and last cutoff scores, a preference-based measure derived from average ranking positions in student applications, and a student quality measure based on the academic credentials of enrolled students. This approach allows the model to distinguish between different aspects of institutional prestige and selectivity that may differentially affect student choices.

We estimate the parameters by Maximum Likelihood and rank programs by the estimated college quality $\hat{\theta}_c$. The implied rankings are reported in Table~\ref{tab:revealed_preference}.\footnote{We normalize one program's $\theta_c$ to zero for identification. Our analysis excludes out-of-pocket tuition costs due to data unavailability. This omission may be less consequential in our setting, as the private medical education sector was considerably smaller during our study period than it has been in recent years.}

\input{tables/revealed_preference_rankings}

The revealed preference rankings are not only inaccurate but also highly unstable across years, which is itself a significant shortcoming: program quality does not change year to year (see Tables \ref{tab:summary_statistics} and \ref{tab:match_outcomes}), and a reliable ranking should reflect this stability. Catholic University's medical program (ID 1258), which consistently ranks first in the desirable rankings and aligns with the contemporary QS World University Rankings, fluctuates between 4th and 20th place in the revealed preference rankings. Conversely, the University of La Frontera's medical program (ID 3026) frequently appears at the top of the revealed-preference rankings, despite ranking seventh in the desirable rankings and occupying a middle-tier position in current national assessments. The University of Valpara\'iso's San Felipe program (ID 1939) moves from fifth place in 2006 to first in 2007, then to twelfth in 2008, and twenty-second in 2009. North Catholic University's program (ID 1887) oscillates between second and fifteenth place. In our setting, the desirable ranking outperforms the revealed preference approach and does not impose the strong data and modeling demands that the latter imposes.

\subsection{Borda Rankings}
The Borda Count is another way to aggregate students' preferences, which ``awards'' points to programs based on their (relative) rank in the students' ROLs.\footnote{For applications of Borda rankings see \cite{Baharad2018} and \cite{Barbera2023}.} To compute the Borda score, we use the procedure outlined in \cite{Emerson2013}: if a student ranks $m$ colleges, then we give $m$ points to their first choice, $(m-1)$ points to their second choice, and so forth. The final ranking is determined by summing these individual points for each program across all students. 

The resulting rankings of medical programs are shown in Table \ref{tab:borda_chile}. The Borda Count ranked the University of Chile (ID 1183) first across all four years from 2006 to 2009, followed by various other institutions in subsequent positions.

\input{tables/borda_rankings}

However, the limitations of Borda's rule become apparent when comparing Borda rankings with contemporary evidence of institutional quality, particularly QS rankings. The most striking divergence involves Chile's two premier medical institutions. While desirable rankings correctly identified Catholic University (ID 1258) as the top program from 2006 to 2009 (consistent with QS rankings), the Borda Count systematically undervalued this elite institution, ranking it third to fifth. This feature suggests that the Borda Count may not reliably distinguish between genuinely high-quality programs and those that appear frequently in student preferences.

One reason for this bias is that Borda's rule systematically favors programs with larger applicant pools, regardless of their quality, because it rewards programs for appearing on more ROLs. Large programs benefit disproportionately because applicants likely include them as safety options, either because the system itself limits the number of programs they can list (and thus they do not want to ``waste'' a slot on a school they might prefer, but think they will be rejected from), or simply because agents choose not to rank alternatives they think are unachievable (see footnote \ref{fn:partial-preferences} and the references therein). This feature will inflate the Borda scores of these programs, leading to incorrect rankings. While this size-induced bias may be less pronounced in the Chilean medical education context, where programs tend to be relatively similar in size, the issue remains problematic, highlighting a fundamental limitation of the Borda approach.

\section{Conclusion}
We consider the problem of ranking alternatives, such as medical programs, colleges, or academic journals. We introduce a new paradigm for constructing rankings that is based on the notion of desirability: alternatives that an agent desires (relative to what she receives) should be ranked higher. 

We introduce axioms that formalize the notion of desirability, and build an algorithm, the Iterated Removal of Underdemand Schools (IRUS), that can be used to calculate desirable rankings. Furthermore, we show that the desirable rankings as an output of the IRUS algorithm coincide with the true underlying rankings based on college qualities in the limit. 

We verify this theoretical convergence through simulations that demonstrate the robust performance of the IRUS algorithm across different market sizes. In our empirical application, we apply IRUS to rank Chilean medical programs from 2006 to 2009 using data only on rank-order lists and admissions outcomes. 
Despite the parsimonious data, our desirable rankings align remarkably well with contemporary program standings. 
We also find that desirable rankings outperform revealed preference rankings and Borda rankings as alternatives.

Having introduced desirability as a ranking paradigm and exhibited its potential in a real-world application, there are several additional conceptual reasons to prefer desirability-based methods, which open up avenues worth further exploration. Desirable rankings will be more robust to idiosyncratic preferences that are unique to a particular student, rather than the underlying quality of the college. By basing the ranking on the colleges a student desires (looking ``up'' a student's preference list, rather than ``down''), we are less likely to make incorrect inferences when a student prefers a lower-quality college for purely idiosyncratic reasons. This robustness is driven by the fact that desirable rankings work by sorting colleges using a shadow matching, which serves to filter out idiosyncratic preferences. This is an important feature of desirable rankings that stands in contrast to counting-based methods (such as the Borda count), which reward schools for simply appearing more frequently and higher on student lists. We expect that such approaches will be more susceptible to bias than desirability if agents inflate the rankings of certain schools for reasons other than quality. Second, as discussed in the Introduction, standard ``index'' methods for calculating rankings (based on factors such as acceptance rates) are problematic because they are susceptible to gaming on behalf of colleges. Notice that under a desirable ranking, for a college to be ranked higher, it must be desired by a student at a lower-ranked school. Thus, if a college wants to raise its rank, it must increase its quality to become a more attractive college. We expect desirable rankings to provide incentives for colleges to invest in quality improvement, rather than strategies that artificially inflate their rankings. Formalizing these properties of desirable rankings are interesting questions for future work.

\bibliographystyle{econ}
\bibliography{master_bib_file}

\label{last-page-main-text}

\appendix

\section{Proofs}

\subsection{Proofs for Section \ref{sec:computing-desirable-rankings}}

\textbf{Proof of Theorem \ref{thm:RUS-is-desirable}.} Let $\bt$ be the ranking produced by the IRUS algorithm using shadow matching $\mu^*$, and let $\Pi=\{\Pi_1,\ldots,\Pi_K\}$ be the corresponding tiers. First, we show that $\bt$ is desirable by showing it satisfies AoD and justification with respect to $\mu^*$. Consider any student $i$ and any college $c$ that $i$ desires. In particular, $c \tp_i \mus(i)$.  Note that $c$ cannot be an underdemanded college until $\mus(i)$ is removed. By construction of the algorithm, $c$ is thus ranked higher than $\mus(i)$, and so $\bt$ satisfies AoD.  

For justification, for any $k<K$ and any school $c\in \Pi_k$, we need to find a sequence of colleges $c_k:=c \sbt c_{k+1}\sbt \cdots \sbt c_K$ and students $i_{k+1},\ldots, i_K$ such that $\mu^*(i_j)=c_j$ and $i_j$ desires $c_{j-1}$ for all $j=k+1,\ldots,K$. We proceed by induction, starting with $k=K-1$.\footnote{Note that we start with $K-1$ because the schools in $\Pi_K$ are ranked last, and do not need to be justified.} Consider $c \in \Pi_{K-1}$.  School $c$ was not removed in step 1 of IRUS, and so in step 1, there is some student $i$ such that $c \tp_{i} \mu^*(i)$. As $c$ is removed in step $2$, it is underdemanded at this step, and so $\mu^*(i)$ must have been removed in step 1, i.e., $\mu^*(i)\in \Pi_K$. Then, $c_{K-1}:=c \sbt c_K:=\mu^*(i)$ and $i_K:=i$ gives the necessary chain to justify $c$. 

 The inductive step is similar. Suppose that the colleges in $\Pi_{k+1}$ have been justified: that is, for any $c_{k+1} \in \Pi_{k+1}$, there exist sequences of colleges $c_{k+1}\sbt c_{k+2}\sbt\cdots \sbt c_K$ and students $i_{k+2}, \ldots, i_{K}$ such that $\mu^*(i_j)=c_j$ and $i_j$ desires  $c_{j-1}$ for all $j=k+2,\ldots,K$. We show that the colleges in $\Pi_k$ are also justified. Consider a college $c\in \Pi_k$, which means that $c$ was removed at step $K-k+1$ of the IRUS algorithm. This means that $c$ was not underdemanded at any step up to and including $K-k$, but is so at step $K-k+1$. Similar to the base case, this means that there is some student $i$ such that $c \tp_{i} \mu^*(i)$ and $\mu^*(i) \in \Pi_{k+1}$. Let $c_{k+1}=\mus(i)$.  By the inductive hypothesis, there exists a sequence of colleges $c_{k+1}\sbt c_{k+2}\sbt \cdots \sbt c_K$ and students $i_{k+2},\ldots,i_{K}$ such that $\mu^*(i_j)=c_j$ and $i_j$ desires $c_{j-1}$ for all $j=k+2,\ldots,K$. Appending school $c_k:=c$ and $i_{k+1}:=i$ to these sequences, we create sequences $c_k:=c\sbt c_{k+1}\sbt c_{k+2}\sbt \cdots \sbt c_K$  and $i_{k+1}:=i,i_{k+2},\ldots,i_{K}$ such that $\mu^*(i_j)=c_j$ and $i_j$ desires $c_{j-1}$ for all $j=k+1,\ldots,K$. So, any school $c\in \Pi_k$ is justified, and by induction, the entire ranking is justified.

To prove uniqueness, let $\bt'$ be a desirable ranking, and $\Pi'=\{  \Pi_1'\ldots,\Pi'_L \}$ be the associated tiers. We show that $\bt=\bt'$ by showing $\Pi_{K-k}=\Pi'_{L-k}$ for all $k=0,1,\ldots, K-1$. Consider $k=0$. We first show that $\Pi_K \subseteq \Pi_L'$. Take some $c\in \Pi_K$. By definition of the IRUS algorithm, college $c$ is underdemanded when the entire set of colleges and students is considered, i.e., $\mu^*(i) \tr_i c$ for all $i\in I$. Therefore, there cannot be a sequence that justifies ranking $c$ in anything but the last tier, as no student desires $c$. Since $\Pi'$ is justified, we conclude that $c\in \Pi_L$ and that $\Pi_K \subseteq \Pi_L'$.  Next, we show $\Pi_L'\subseteq \Pi_K$. Take some $c\notin \Pi_K$. Since $c\notin \Pi_K$, it is not underdemanded at step 1 of IRUS, and so there must be some student $i\in I$ such that $c \tp_i \mu^*(i)$. If $c\in \Pi_L'$, then $\mu^*(i)\bt c$ (as $c$ is ranked last, any school is ranked weakly higher than it). But, $c \tp_i \mu^*(i)$ and $\mu^*(i)\bt c$ violates the axiom of desire. Therefore, $c\notin \Pi_L'$, $\Pi_L'\subseteq \Pi_K$, and $\Pi_K=\Pi_L'$.

The inductive step is analogous. Suppose that $\Pi_{K-k'}=\Pi_{L-k'}'$ for all $k'=0,\ldots,k-1$. Let $X=\Pi_{K-k+1}\cup \Pi_{K-k+2}\cup \ldots \cup \Pi_K$ be the set of all colleges ranked $K--k+1$ or worse according to IRUS, and let $\bar{X}=C\setminus X = \Pi_1\cup\Pi_2 \cup\cdots \cup \Pi_{K-k}$ be the set of schools ranked $K-k$ or better. Note that by the inductive hypothesis, under the alternative ranking $\bt'$, we have $X=\Pi'_{L-k+1}\cup  \ldots \cup\Pi_L'$. We show that $\Pi_{K-k}=\Pi'_{L-k}$. 

We first show $\Pi_{K-k}\subseteq \Pi_{L-k}'$. Take some $c\in \Pi_{K-k}$.   At the step of IRUS at which $c$ was removed, the set of remaining colleges is exactly $\bar{X}$. Let $\bar{I}=\{i\in I: \mu^*(i)\in \bar{X}\}$ be the students at these colleges. When $c$ was removed, it was underdemanded, and so $\mu^*(i) \tr_i c$ for all $i\in \bar{I}$. By the inductive hypothesis, $c \notin \Pi_{\ell}'$ for $\ell\geq K-k+1$. We claim that $c\notin \Pi_\ell'$ for $\ell<K-k$, which will imply that $c\in \Pi_{K-k}'$. Indeed, assume that $c\in \Pi_\ell'$ for some $\ell<K-k$. Since the ranking is justified, there exist sequences of colleges $c_\ell:=c\sbt c_{\ell+1} \sbt \cdots \sbt c_{K-k+1}\sbt \cdots \sbt c_K$ and students $i_{\ell+1},\ldots,i_K$ such that $\mu^*(i_{\ell'})=c_{\ell'}$ and $i_{\ell'}$ desires $c_{\ell'-1}$ for all $\ell'=\ell+1,\ldots,K$. In particular, to construct such sequences, there must be some student $i\in \bar{I}$ that desires $c$, i.e. $c \tp_i \mu^*(i)$ for some $i\in \bar{I}$, which is a contradiction. Therefore, $c\in \Pi_{L-k}'$ and $\Pi_{K-k}\subseteq \Pi_{L-k}'$. 

Finally, we show that $\Pi_{L-k}'\subseteq \Pi_{K-k}$. Take a college $c\notin \Pi_{K-k}$.  We show that $c\notin \Pi_{L-k}'$. If $c\in \Pi_\ell$ for $\ell>K-k$, then by the inductive hypothesis, $c\in \Pi_\ell'$, and thus $c\notin \Pi_{L-k}'$. So, consider $c\in \Pi_\ell$ for $\ell < K-k$. Since $c\notin \Pi_{K-k}$, during the step of IRUS when the colleges $\Pi_{K-k}$ were removed, college $c$ was still present and not removed, and thus, was not underdemanded, which implies that $c \tp_i \mu^*(i)$ for some $i\in \bar{I}$. By the inductive hypothesis, school $\mu^*(i) \notin \Pi_{L-k+1}'\cup \cdots \cup \Pi_L'$, i.e., school $\mu^*(i)$ must be ranked $L-k$ or better under $\bt'$. If $c\in \Pi_{L-k}'$, then $\mu^*(i) \bt c$ and $c \tp_i \mu^*(i)$, which violates the axiom of desire. Therefore, $c \notin \Pi_{L-k}'$, and $\Pi_{K-k}=\Pi_{L-k}'$.

\subsection{Proofs for Section \ref{sec:convergence}\label{sec:convergence_proof}}

The proofs for this section are presented out of order. We use Theorem \ref{thm:small-tiers} to prove \ref{thm:correct-rankings-limit}, and so we begin by proving Theorem \ref{thm:small-tiers}, followed by Corollary \ref{lem:rankings-convergence}, and finally Theorem \ref{thm:correct-rankings-limit}. \bigskip

\noindent\textbf{Proof of Theorem \ref{thm:small-tiers}.} For each market realization, let $\bt$ be a desirable ranking , and let $X$  denote the largest tier of schools, i.e., $X=\Pi^\bt_k$ for the tier $k$ such that $|\Pi_k^\bt|\geq |\Pi_{k'}^\bt|$ for all $k'$.\footnote{If there are multiple largest tiers, we pick one of them.}     An equivalent way to state Theorem \ref{thm:correct-rankings-limit} is as follows: For any $\tau,\lambda > 0$, there exists $N$ such that
$$
\text{Pr}\left( \frac{1}{n}|X| > \tau \right) < \lambda \text{ for every }n>N.
$$
This is the statement that we will show. 

Since $\bt$ is desirable, it satisfies the axiom of desire with respect to a particular shadow matching, which we denote $\mu^*$. As a reminder, $\mu^*$ is Pareto-efficient by definition. Let $I_X = \{i \in I | \mus(i)\in X\}$ be the students assigned to the colleges in $X$ at $\mu^*$.  

Let $\Delta \in (0,\min\{1, \frac{1-\alpha}{\alpha}\})$. Let $p_\Delta = 1-\frac{\alpha}{1-\alpha}\Delta$, and notice that $p_\Delta \in (0,1)$. Define a bipartite graph $G^{p_\Delta}$ as follows: $X\cup I_X$ is a bipartitioned set of nodes. Two vertices $c\in X$ and $i\in I_X$ are joined by an edge if and only if 
\begin{equation}\label{eqn:edge-condition}
\eta_{i,c}\leq p_\Delta.
\end{equation}
As the $\eta_{i,c}$'s are iid uniform[0,1], every student and college is joined by an edge independently with probability $p_\Delta$, and so indeed this fits into the bipartite random graph model. 

There is a connection between a student $i$ having unusual preferences for a school $c$ (in the sense that $i$ prefers a college with a lower common value) and there being an edge between $i$ and $c$ in our random graph $G^{p_\Delta}$.

\begin{lemma}\label{lem:unusal-students}

For any student $i\in I_X$ and college $c\in X$, if there exists a college $d\in X$ such that (i) $\theta_c > \theta_d+\Delta$ and (ii) $d \tp_i c$, then there is an edge between $i$ and $c$ in $G^{p_\Delta}$.

\end{lemma}

\noindent\textbf{Proof of Lemma \ref{lem:unusal-students}.}
Since $d \tp_i c$, we have the following string of inequalities:
\begin{align*}
\alpha \theta_d + (1-\alpha)\eta_{i,d} &> \alpha \theta_c + (1-\alpha)\eta_{i,c}\\
\alpha \theta_d +(1-\alpha) &> \alpha \theta_c + (1-\alpha)\eta_{i,c}\\
\alpha(\theta_d - \theta_c)+(1-\alpha) &>  (1-\alpha)\eta_{i,c}\\
1 - \frac{\alpha}{1-\alpha}(\theta_c - \theta_d) &>  \eta_{i,c}\\
1- \frac{\alpha}{1-\alpha}\Delta &>  \eta_{i,c}\\
p_\Delta &> \eta_{i,c}.\end{align*}
The second line follows from $\eta_{i,d}<1$.  The second-to-last line follows from the assumption that $\theta_c-\theta_d>\Delta$. The remaining lines are algebra. So, indeed, given the construction of $G^{p_\Delta}$ (see the edge condition given by Equation \ref{eqn:edge-condition}), there is an edge between $i$ and $c$.

\hfill $\square$

A consequence of Lemma \ref{lem:unusal-students}, is that a large tier (greater than $\Delta$\% of the market) induces a biclique in the random graph $G^{p_\Delta}$. To state this formally, given two numbers $0\leq a < b \leq 1$, define
\begin{eqnarray*}
L(a) &=& \{ c\in X| \theta_c\leq a\}\\
H(b) &=& \{c\in X| \theta_c\geq b\}
\end{eqnarray*}
to be the subsets of colleges with low (below $a$) and high (above $b$) common values, respectively. Recall that $\Delta \in (0,\min\{1, \frac{1-\alpha}{\alpha}\})$.

\begin{lemma}\label{lem:biclique} For any $a,b\in (0,1)$ such that $\Delta = b-a$, the graph $G^{p_\Delta}$ has a balanced biclique of size $min\{|L(a)|,|H(b)|\}.$
\end{lemma}

\noindent\textbf{Proof of Lemma \ref{lem:biclique}.} Since $\bt$ is desirable,  for any student $i\in I_X$, $\mus(i)$ is $i$'s top-ranked school in $X$: $\fav_i(X)=\mu^*(i)$.\footnote{\label{fn:aod-implies-fave-college}By the axiom of desire, $c' \sbt \mu^*(i)$ for all $c'$ such that $c' \tp_i \mu^*(i)$, and thus any college $i$ strictly prefers to $\mu^*(i)$ is ranked in a tier strictly better than $X$.} Let
$$U = \left\{ i\in X: \mus(i)\in L(a) \right\}.$$
We use the notation $U$ because these students are ``unusual'', in that their favorite tier $X$ college has relatively low common value (below $a$).

\begin{claim}
In the graph $G^{p_\Delta}$, $H(b)$ and $U$ are a biclique.
\end{claim}

\noindent\textbf{Proof of claim.} Choose a student $i\in U$ and college $c\in H(b)$.  Let $d=\mus(i)$.  Since $i\in U$, $d \in L(a)$.  Therefore, $\theta_d <a$.  Since $c\in H(b)$, $\theta_c >b$.  Therefore, $\theta_c-\theta_d>\Delta$.  As $d=\fav_i(X)$ we have $d \tp_i c$.  Therefore, by Lemma \ref{lem:unusal-students}, there is an edge in $G^{p_\Delta}$ between $i$ and $c$. As this holds for each choice of $i\in U$ and $c\in H(b)$, there is an edge between every $i\in U$ and $c\in H(b)$, and thus, $H(b)$ and $U$ are a biclique. 

\hfill $\square$

Since $H(b)$ and $U$ form a biclique, there exists a \emph{balanced} biclique of size $\min\{|U|,|H(b)|\}$.  For each college $c\in L(a)$, choose one student $i_c\in U$.  Let $V=\{  i_c \in U : c\in L(a) \}$ be this set.  Since $V\subset U$, $V$ and $H(b)$ form a biclique.  Since $|V|=|L(a)|$,  there is a balanced biclique of size $B\times B$, where $B=\min\{|L(a)|,|H(b)|\}$.

\hfill $\square$

\begin{lemma}\label{lem:large-tier-implies-biclique}
For every $\tau, \delta>0$, there exists $\epsilon>0$, $\Delta \in (0,\min\{1, \frac{1-\alpha}{\alpha}\})$ and $N> 0$ such that for all $n>N$, 
$$
\text{Pr}\left( \frac{1}{n}|X| > \tau \right) \leq \text{Pr}\left(\frac{B}{n}> \epsilon \right) + \delta,
$$
where $B\times B$ is the size of the maximal balanced biclique in the associated graph $G^{p_\Delta}$.
\end{lemma}

\noindent\textbf{Proof of Lemma \ref{lem:large-tier-implies-biclique}.}  Fix any $\tau\in (0,1)$,\footnote{If $\tau\geq 1$, then the LHS of in the inequality in the lemma is 0, and the result holds trivially.} and let $M$ be an integer such that $\frac{1}{M}<\min\{\frac{\tau}{4},\frac{1-\alpha}{\alpha}\}$. Choose $\epsilon>0$ such that $\epsilon <\frac{\tau}{4M}$, and set $\Delta = 1/M$. (Notice that $\Delta<\min\{\frac{\tau}{4},\frac{1-\alpha}{\alpha}\}<\min\{1,\frac{1-\alpha}{\alpha}\}$, which was the condition we needed for the lemmas above.) Divide the unit interval into $M$ subintervals, each with length $1/M$, where we refer to $\left[\frac{m-1}{M},\frac{m}{M}\right)$ as the $m^{th}$ \textbf{subinterval}. Let 
$$
C^m=\left\{ c\in C: \frac{m}{M} \leq \theta_c < \frac{m+1}{M} \right\}
$$
be the colleges with common values in the $m^{th}$ subinterval. Let $E_n$ be the event that $\frac{1}{n}|X| > \tau$, and $F_n$ the event that $\frac{|C^m|}{n} < \frac{\tau}{4} \text{ for all subintervals } m=1,\ldots,M$.
\begin{claim}\label{clm:biclique-if-F-true}
Assume that the events $E_n$ and $F_n$ are true. Then, in the graph $G^{p_\Delta}$, there is a balanced biclique of size $B\times B$, where $B = \lceil \epsilon n \rceil$. 
\end{claim}

\noindent\textbf{Proof of Claim \ref{clm:biclique-if-F-true}.} Since $F_n$ is true, we have $|C^m|<(\tau/4)\times n$ for all $m$, which also implies that $|C^m \cap X|<(\tau/4)\times n$ for all $m$. The event $E_n$ is $\frac{1}{n}|X| > \tau$, and a necessary condition for this to be true is that there is some $m^{th}$ subinterval that contains the common values of at least $\epsilon \times n$ of the colleges in $X$.\footnote{Indeed, if not, then $|X| < M\times\epsilon \times n<M \times \tau/(4M)\times n < \tau \times n$, which implies $|X|<\tau n$.} Let $m$ be the index of the first such subinterval that contains the common values of at least $\epsilon \times n$ of the colleges in $X$. Combining these observations, we have:
\begin{eqnarray}\label{eqn:first-m+1-intervals}
\left(\sum_{m'=1}^{m-1} |C^{m'} \cap X| \right)+|C^{m} \cap X|+|C^{m+1} \cap X|&\leq&  (m-1)\epsilon n + \frac{\tau}{4}n + \frac{\tau}{4}n \nonumber \\
&<& (m-1)\frac{\tau}{4M}n + \frac{\tau}{4}n + \frac{\tau}{4}n \nonumber \\
&=& \frac{m-1}{M}\frac{\tau}{4}n +\frac{\tau}{4}n + \frac{\tau}{4}n \nonumber \\
&<&\frac{\tau}{4}n +\frac{\tau}{4}n + \frac{\tau}{4}n\nonumber\\ &=& \frac{3}{4}\tau n.
\end{eqnarray}

The first inequality follows from the fact that each of the first $m-1$ intervals has less than $\epsilon n$ of the colleges in $X$ (by definition of $m$) and intervals $m$ and $m+1$ have at most $(\tau/4)n$ total colleges (and therefore, must have less than that number of $X$ colleges). The second inequality comes from the fact that we chose $\epsilon < \frac{\tau}{4M}$. The remaining inequalities are simple algebra.

Note that $\sum_{m'=1}^{M} |C^{m'} \cap X| = |X| > \tau n$. Combining this with Equation \ref{eqn:first-m+1-intervals} implies that 
\begin{eqnarray*}
\sum_{m'=m+2}^M |C^{m'} \cap X| &>& \frac{\tau n}{4} >\epsilon n,
\end{eqnarray*}
which follows because we chose $\epsilon < \tau/4$. In sum, we have shown that there are at least $\epsilon n$ colleges in $X$ with common values in the interval $\left[0, \frac{m+1}{M} \right)$, and at least $\epsilon n$ colleges with common values in the interval $\left[\frac{m+2}{M},1\right]$. Therefore, by Lemma \ref{lem:biclique}, setting $a=\frac{m+1}{M}$, $b=\frac{m+2}{M}$, and $\Delta=\frac{1}{M}$,\footnote{Recall that we chose $M$ large enough such that $1/M<\min\{\tau/4,(1-\alpha)/\alpha\}$, and so the conditions of Lemma \ref{lem:biclique} are satisfied.} we conclude that in the random graph $G^{p_\Delta}$, there is a balanced biclique of size at least $\epsilon n$.

\hfill $\square$

\noindent Recall that $E_n$ is the event that $\frac{1}{n}|X| > \tau$, and $F_n$ is the event that $\frac{|C^m|}{n} < \frac{\tau}{4}$  for all subintervals  $m=1,\ldots,M$.  Our goal is to show that $\text{Pr}(E_n) \leq \text{Pr}(B/n > \epsilon) +\delta$. Note that we can write:
\begin{eqnarray}\label{eqn:events}
\text{Pr}(E_n) &=& \text{Pr}(E_n\cap F_n) + \text{Pr}(E_n\cap F_n^c) \nonumber \\
&\leq & \text{Pr}(E_n\cap F_n) + \text{Pr}(F_n^c),
\end{eqnarray}
where $F_n^c$ is the complement of $F_n$, i.e., $F^c_n$ is the event that there is some subinterval $m'$ such that $|C^{m'}|/n > \tau/4$. Because all $\theta_c$ are iid uniform, $\frac{|C^m|}{n}\stackrel{p}{\rightarrow}\frac{1}{M}$ for each $m$, and thus for any $\epsilon',\delta'>0$, for all $n$ large enough,
\begin{eqnarray}\label{eqn:all-intervals-1/M-draws}
\text{Pr}\left(\left\{\left| \frac{|C^m|}{n}-\frac{1}{M}\right|<\epsilon' \text{ for all } m\right\}\right)>1-\delta'.
\end{eqnarray}

Further, because $1/M < \tau/4$, this also implies that for all $n$ large enough, 
\begin{eqnarray}\label{eqn:all-intervals-less-than-Delta/4}
\text{Pr}\left( F_n \right)>1-\delta'.
\end{eqnarray}
for all $\delta'>0$. In other words, with arbitrarily high probability, for all $n$ large enough, every interval contains at most $\frac{\tau}{4}\times n$ of the colleges. This can be restated as 
\begin{eqnarray}\label{eqn:Fc-small}
\text{Pr}\left( F_n^c \right) \leq \delta' \text{ for all sufficiently large $n$}.
\end{eqnarray}
Combining equations (\ref{eqn:events}) and (\ref{eqn:Fc-small}), and choosing $\delta'=\delta$, we have 
\begin{eqnarray}\label{eqn:E-less-than-EF-plus-delta}
\text{Pr}(E_n) \leq \text{Pr}(E_n\cap F_n) + \delta \text{ for all $n$ sufficiently large.}
\end{eqnarray}
By Claim \ref{clm:biclique-if-F-true}, $\text{Pr}(E_n\cap F_n) \leq \text{Pr}(B/n > \epsilon)$, and so 
\begin{eqnarray}
\text{Pr}(E_n) \leq \text{Pr}(B/n > \epsilon) +\delta \text{ for all $n$ sufficiently large,}
\end{eqnarray}
which is what we wanted to show.

\hfill $\square$

To finish the proof of Theorem \ref{thm:small-tiers}, let $\beta_n=2\log(n)/\log(1/(p_\Delta))$, where $\Delta$ is chosen as in Lemma \ref{lem:large-tier-implies-biclique}, and let $B\times B$ be the size of the maximal balanced biclique in the graph $G^{p_\Delta}$. The theorem of \cite{dawande2001bipartite} implies that 
$$
\text{Pr}(B\leq \beta_n) \rightarrow 1 \text{ as } n\rightarrow \infty
$$

and, since $\beta_n/n \rightarrow 0$, we have 
\begin{eqnarray}\label{eqn:K/n-goes-to-0}
\frac{B}{n} \stackrel{p}{\longrightarrow} 0 \text{ as }n\rightarrow \infty.
\end{eqnarray}
Now, by Lemma \ref{lem:large-tier-implies-biclique}, there exists $N$ such that for all $n>N$, we have
\begin{eqnarray}\label{eqn:lemma-3-restated}
\text{Pr}\left( \frac{1}{n}|X| > \tau \right) \leq \text{Pr}\left(\frac{B}{n}> \epsilon \right) + \delta
\end{eqnarray}
By (\ref{eqn:K/n-goes-to-0}), the first term on the RHS of (\ref{eqn:lemma-3-restated}) converges to 0, and thus by choosing $\delta$ sufficiently small and $n$ sufficiently large, we can make the RHS of (\ref{eqn:lemma-3-restated}) smaller than any $\lambda>0$, which completes the proof of Theorem \ref{thm:small-tiers}.

\hfill $\square$

\textbf{Proof of Corollary \ref{lem:rankings-convergence}.} Fix $\tilde{\theta}\in [0,1]$. We first show that $|D(\tilde{\theta})|/n \stackrel{p}{\rightarrow}1-\tilde{\theta}$. First notice that $|D(\tilde{\theta})|/n\leq 1-\tilde{\theta}$, which follows simply from the definition of $\rho(\cdot)$.  However, the converse does not immediately follow, i.e., it may be that $|D(\tilde{\theta})|/n>1-\tilde{\theta}$. We show that as $n$ grows large, $|D(\tilde{\theta})|/n$ becomes arbitrarily close to $1-\tilde{\theta}$ with high probability, i.e., we show that, for any $\epsilon>0$, 
 \begin{eqnarray}\label{eqn:D-converges-1-theta}
\text{Pr}\left( \frac{|D(\tilde{\theta})|}{n} > (1-\tilde{\theta})-\epsilon\right)\rightarrow 1 \text{ as } n\rightarrow\infty.
 \end{eqnarray}
Define the following quantities:
 \begin{eqnarray*}
 \underline{\rho} := \max_{c\in C} \{\rho(c) : \rho(c)<\tilde{\theta}\} \\
 \bar{\rho} := \min_{c\in C} \{\rho(c) : \rho(c) \geq \tilde{\theta}\}.
 \end{eqnarray*}
By construction, $\bar{\rho}\geq \tilde{\theta}\geq \underline{\rho}$ (notice that these inequalities may be strict). Also, by the definition of $\rho(\cdot)$, we have 
$$
\bar{\rho} - \underline{\rho} = \frac{|\{c\in C : \rho(c) =\underline{\rho}\}|}{n}.
$$
the numerator on the RHS is a single ranking tier, and so becomes small as $n$ grows large; formally, by Theorem \ref{thm:small-tiers}, the RHS of the above equation converges in probability to 0, and thus $\bar{\rho}  \stackrel{p}{\rightarrow} \underline{\rho}$. Combined with $\bar{\rho}\geq \tilde{\theta}\geq \underline{\rho}$, we have
\begin{eqnarray}\label{eqn:rhobar-to-thetatilde}
\bar{\rho}\stackrel{p}{\rightarrow} \tilde{\theta}.
\end{eqnarray}
Define
$$
E = \{c\in C : \rho(c)\geq \bar{\rho}\}.
$$
Since $\bar{\rho}\geq \tilde{\theta}$, we have $E \subseteq D(\tilde{\theta})$, which implies that 
\begin{eqnarray}\label{eqn:D-bigger-than-E}
\text{Pr}\left( \frac{|D(\tilde{\theta})|}{n} > (1-\tilde{\theta})-\epsilon\right) \geq \text{Pr}\left( \frac{|E|}{n} > (1-\tilde{\theta})-\epsilon\right)
\end{eqnarray}
for any $\epsilon >0$. Next, notice that by definition of $\bar{\rho}$, $|E|/n=1-\bar{\rho}$, so the RHS of (\ref{eqn:D-bigger-than-E}) becomes 
$$
\text{Pr}(1-\bar{\rho}>(1-\tilde{\theta})-\epsilon)=\text{Pr}(\bar{\rho}<\tilde{\theta}+\epsilon).
$$
By (\ref{eqn:rhobar-to-thetatilde}), $\text{Pr}(\bar{\rho}<\tilde{\theta}+\epsilon) \rightarrow 1$. So, the RHS of (\ref{eqn:D-bigger-than-E}) converges to 1, and thus  
$$
\text{Pr}\left( \frac{|D(\tilde{\theta})|}{n} > (1-\tilde{\theta})-\epsilon\right) \geq \text{Pr}\left( \frac{|E|}{n} > (1-\tilde{\theta})-\epsilon\right)\rightarrow 1,
$$
and thus, 
$$\text{Pr}\left( \frac{|D(\tilde{\theta})|}{n} > (1-\tilde{\theta})-\epsilon\right) \rightarrow 1 \text{ as } n\rightarrow \infty$$ which is what we wanted to show. The second statement that $|D'(\tilde{\theta})|/n\stackrel{p}{\rightarrow} \tilde{\theta}$ is an immediate consequence of the first. 

 \hfill $\square$

\textbf{Proof of Theorem \ref{thm:correct-rankings-limit}.} For each market realization, let $\bt$ be a desirable ranking, and $\rho$ the induced percentage ranking. Since $\bt$ is desirable, it satisfies the axiom of desire with respect to a particular shadow matching, which we denote $\mu^*$. As a reminder, $\mu^*$ is a Pareto-efficient matching. To prove Theorem \ref{thm:correct-rankings-limit}, we first show the following two propositions.

\begin{proposition}\label{loser-schools-go-to-0}
 For every $\epsilon >0$ and any $\theta \in (0,1]$,
$$ 
\frac{1}{n}|\left\{ c\in C : \theta_c \geq \theta \text{ and } \rho(c) \leq \theta - \epsilon \right\}| \xrightarrow{p} 0 \text{ as } n\rightarrow \infty.
$$
\end{proposition} 

\begin{proposition}\label{winner-schools-go-to-0}  For every $\epsilon >0$ and any $\theta \in (0,1]$,
$$ 
\frac{1}{n}|\left\{ c\in C : \theta_c \leq \theta \text{ and } \rho(c) \geq \theta + \epsilon \right\}| \xrightarrow{p} 0 \text{ as } n\rightarrow \infty.
$$
\end{proposition}

\textbf{Proof of Proposition \ref{loser-schools-go-to-0}.} Fix any $\theta \in (0,1]$ and any $\epsilon>0$.  Choose $\theta'$ so that $\theta>\theta'>\theta - \min\{\epsilon, \frac{1-\alpha}{\alpha}\}$.

Define the set of colleges that ``lose'' in ranking $\rho$ by
$$
L:= \left\{ c \in C : \theta_c  \geq \theta \text{ and }\rho(c)\leq\theta-\epsilon \right\}.
$$
These colleges ``lose'' because their quality $\theta_c$ is above $\theta$, but their resulting ranking is more than $\epsilon$ below. Next, define the following set of colleges:
$$
W:= \left\{ c \in C : \theta_c  < \theta' \text{ and }\rho(c)>  \theta-\epsilon \right\}.
$$
We label this set $W$ because these colleges are ``wrongly ranked'', in the sense that their true quality $\theta_c$ is less than all colleges in $L$, and yet they are ranked higher than any of the colleges in $L$.
As the shadow matching $\mus(i)$ is Pareto efficient, $\mus(i)$ is $i$'s top-ranked college that is in the same tier as $\mus(i)$ (see footnote \ref{fn:aod-implies-fave-college}). We define a set of ``unusual'' students as follows:
$$
U:= \left\{ i \in I:\mus(i) \in W \right\}.
$$
These students are unusual in the sense that the colleges in $W$ are overranked, and thus have relatively lower common values, and yet each of these students has a college in $W$ as her favorite college in her tier.  

\begin{claim}\label{clm:unusual-students-don't-prefer-loser-colleges}
For every college $c\in L$ and every student $i\in U$,  $\mus(i) \tp_i c$.
\end{claim}   

This claim follows from definitions of $W$, $L$, and AoD.  Since $i\in U$, $\mus(i) \in W$.  Therefore, $\rho(\mus(i))>\rho(c)$ (from the definitions of $W$ and $L$), and by AoD, $\mus(i) \tp_i c$.

In fact, we can place an upper bound on the idiosyncratic draw between an unusual student and a loser college.\footnote{\label{fn:theta-theta'-bigger-zero}Recall that $\theta'$ is chosen such that $\theta'>\theta-\min\{\epsilon,\frac{1-\alpha}{\alpha}\}$, which implies that $1-\frac{\alpha}{1-\alpha}(\theta-\theta')>0$.}

\begin{claim}\label{clm:upper-bound-on-unus-stu-draws}
For every college $c\in L$ and every student $i\in U$, $\eta_{i,c} < 1-\frac{\alpha}{1-\alpha}(\theta-\theta')$.
\end{claim}

To see Claim \ref{clm:upper-bound-on-unus-stu-draws}, recall that $\theta_c\geq\theta$ (by the definition of $L$).  By definition of $U$, $\mus(i)\in W$ and therefore, $\theta' > \theta_{\mus(i)}$, by definition of $W$.  Since $\theta'> \theta_{\mus(i)}$ and $1\geq\eta_{i,\mus(i)}$ 
\begin{align*}
\alpha\theta'+(1-\alpha) &> \alpha \theta_{\mus(i)}+(1-\alpha)\eta_{i,\mus(i)} \\
&> \alpha \theta_c + (1-\alpha)\eta_{i,c}\\
&\geq\alpha\theta + (1-\alpha)\eta_{i,c}
\end{align*}

The second line follows since $\mus(i) \tp_i c$ (Claim \ref{clm:unusual-students-don't-prefer-loser-colleges}).  The last line follows by construction:  All colleges in $L$ have a common value greater than or equal to $\theta$. Thus, 
$$\alpha \theta'+(1-\alpha) > \alpha \theta +(1-\alpha)\eta_{i,c},$$
and rearranging terms, we find that
$$1-\frac{\alpha}{1-\alpha}(\theta-\theta') > \eta_{i,c},$$
which is Claim \ref{clm:upper-bound-on-unus-stu-draws}.

We now use the theory of random bipartite graphs, and show that our ``unusual'' occurrence corresponds to the formation of a balanced biclique in a random graph. Let $I \cup C$ be a bipartitioned set of nodes. Construct a graph by drawing an edge between student $i$ and college $c$ if 
\begin{eqnarray}\label{eq:edge-condition}
\eta_{i,c} \leq 1- \frac{\alpha}{1-\alpha}(\theta-\theta').
\end{eqnarray}
That is, every pair is joined by an edge independently with probability $p'=1-\frac{\alpha}{1-\alpha}(\theta-\theta')$,\footnote{Given how $\theta'$ is chosen, $p'\in(0,1)$; see footnote \ref{fn:theta-theta'-bigger-zero}.}  and we denote the resulting random graph $G^{p'}$. 

\begin{claim}\label{clm:L-U-biclique}
The vertices $L$ and $U$ are a biclique in $G^{p'}$.
\end{claim}  

    This result follows immediately from Claim \ref{clm:upper-bound-on-unus-stu-draws} and equation (\ref{eq:edge-condition}). Thus, in the random graph, there exists a \emph{balanced} biclique of size $\min\{|L|, |U|\}$. 
    
For each college $c\in W$, choose one student $i_c$. Let $V=\{  i_c : c\in W \}$ be this set of students. Since $V\subset U$, $V$ and $L$ form a biclique.  Since $|V|=|W|$,  there is a balanced biclique of size at least $\min\{|L|,|W|\}$.  By the theorem of \cite{dawande2001bipartite},\footnote{The full argument for (\ref{eqn:min-L-W-to-0}) is analogous to that used to show (\ref{eqn:K/n-goes-to-0}) in the proof of Theorem \ref{thm:small-tiers}.}
\begin{eqnarray}\label{eqn:min-L-W-to-0}
\min\left\{\frac{|L|}{n},\frac{|W|}{n}\right\} \stackrel{p}{\rightarrow} 0.
\end{eqnarray} 

Finally, we have the following claim. 
\begin{claim}\label{claim:winner-schools-converge-positive}
$|W|/n \stackrel{p}{\rightarrow} b>0 $.
\end{claim}

Claim \ref{claim:winner-schools-converge-positive} is sufficient to prove Proposition \ref{loser-schools-go-to-0}, because if $|W|/n$ converges to something strictly positive, then by (\ref{eqn:min-L-W-to-0}), we must have $|L|/n \stackrel{p}{\rightarrow} 0$. To show Claim \ref{claim:winner-schools-converge-positive}, let
$$
A=\{c \in C | \rho(c) \geq \theta - \epsilon\}.
$$
Notice that by definition, $W\subseteq A$. Let $W^c = A \setminus W$, i.e., 
$$
W^c = \{c\in C: \theta_c \geq \theta' \text{ and } \rho(c)\geq \theta-\epsilon \}
$$
Since $A = W \cup W^c$ and $W\cap W^c =\emptyset$, we have 
\begin{eqnarray}\label{eqn:A-W-Wc}
\frac{|A|}{n}= \frac{|W|}{n} + \frac{|W^c|}{n}
\end{eqnarray}
By Corollary \ref{lem:rankings-convergence}, $|A|/n \stackrel{p}\rightarrow 1-(\theta-\epsilon)$.  Define $\omega$ by $|W^c|/n \stackrel{p}{\rightarrow} \omega$, and we have 
\begin{eqnarray}\label{eqn:W/n-plim}
 \frac{|W|}{n} \stackrel{p}{\rightarrow} 1-(\theta-\epsilon) - \omega.
\end{eqnarray}
It follows from the definition of $W^c$ that $\omega\leq 1-\theta'$, and as $\theta'> \theta - \epsilon$, the RHS of (\ref{eqn:W/n-plim}) is strictly positive. 

\hfill $\square$

\textbf{Proof of Proposition \ref{winner-schools-go-to-0}.} The proof of Proposition \ref{winner-schools-go-to-0} proceeds in an analogous manner to the proof of Proposition \ref{loser-schools-go-to-0}. Fix $\theta\in(0,1]$ and $\epsilon>0$. Choose $\theta'$ such that $\theta < \theta'< \theta+\min\{\epsilon, \frac{1-\alpha}{\alpha}\}$. Define the following sets: 
\begin{eqnarray*}
W:=\{c\in C | \theta_c\leq \theta \text{ and } \rho(c)\geq \theta+\epsilon\} \\
L:=\{c\in C | \theta_c > \theta' \text{ and } \rho(c) < \theta + \epsilon\} \\
U:=\{i\in I: \mus(i)\in W\}
\end{eqnarray*}
Arguments analogous to Claims \ref{clm:unusual-students-don't-prefer-loser-colleges}-\ref{clm:L-U-biclique} in the proof of Proposition \ref{loser-schools-go-to-0} deliver the same conclusion as equation (\ref{eqn:min-L-W-to-0}):
$$
\min\left\{\frac{|L|}{n},\frac{|W|}{n}\right\}\stackrel{p}{\rightarrow} 0.
$$
We now show that in this case, it is $|L|/n$ that converges to something strictly positive, which will imply that $|W|/n\stackrel{p}{\rightarrow}0$ and complete the proof.

\begin{claim}
$|L|/n \stackrel{p}{\rightarrow} b > 0.$
\end{claim}
To see this claim, define $A=\{c\in C| \rho(c)< \theta + \epsilon\}$ and $L^c=\{c\in C| \theta_c \leq \theta' \text{ and } \rho(c)<\theta+\epsilon\}$. Notice that $A=L \cup L^c$ and $L\cap L^c = \emptyset$, and therefore,
$$
\frac{|A|}{n} = \frac{|L|}{n} + \frac{|L^c|}{n}. 
$$
By Corollary \ref{lem:rankings-convergence}, $|A|/n \stackrel{p}{\rightarrow}\theta+\epsilon$, and thus, $
\frac{|L|}{n} \stackrel{p}{\rightarrow} \theta+\epsilon - \beta,$
where $\beta \stackrel{p}{\leftarrow}|L^c|/n$. It is clear from the definition of $L^c$ that $\beta\leq \theta'$, and, as $\theta'<\theta+\epsilon$, we have $\theta+\epsilon -\beta>0$, which completes the proof.

\hfill $\square$

Finally, we use Propositions \ref{loser-schools-go-to-0} and \ref{winner-schools-go-to-0} to complete the proof of Theorem \ref{thm:correct-rankings-limit}. Recall that $C^\rho(\epsilon)$ is the set of schools whose percentile ranking, $\rho(c)$, is within $\epsilon$ of their common values, $\theta_c$, i.e., $\theta_c-\epsilon < \rho(c) < \theta_c +\epsilon$. We show that, for any $\epsilon,\delta>0$,
\begin{eqnarray}\label{eq:convergence-non-target-schools}
\text{Pr}\left( \frac{|C\setminus C^\rho(\epsilon)|}{n} > \delta \right) \rightarrow 0 \text{ as } n\rightarrow \infty
\end{eqnarray}
Choose a positive integer $M$ such that $1/M <\min\{\delta, \epsilon/2\}$. Now, we have the following:
\begin{eqnarray}\label{eq:number-nontarget-schools-ineq}
\frac{|C \setminus C^\rho(\epsilon)|}{n} &\leq& \sum_{m=1}^M\frac{1}{n} |\{c\in C : \theta_c \geq m/M \text{ and } \rho(c) \leq m/M - \epsilon/2 \}| + \nonumber \\
&&  \sum_{m=1}^M\frac{1}{n} |\{c\in C : \theta_c \leq m/M \text{ and } \rho(c) \geq m/M +  \epsilon/2 \}| + \nonumber \\ 
&&  \frac{1}{n} |\{c\in C: \theta_c \leq 1/M\}|.
\end{eqnarray}
To see this, take a college $c\in C\setminus C^\rho(\epsilon)$. We claim that $c$ belongs to one of the three terms on the right-hand side of (\ref{eq:number-nontarget-schools-ineq}). If $\theta_c < 1/M$, this is obvious. If not, then there is some $m_c \in \{1,2,\ldots, M-1\}$ such that $\theta_c \in [ m_c/M, (m_c+1)/M ]$. Since $c\notin C^\rho(\epsilon)$, we have that either: (i) $\rho(c)\geq\theta_c+\epsilon$ or (ii) $\rho(c)\leq\theta_c-\epsilon$. If (i) holds, then
\begin{eqnarray*}
\rho(c) \geq \theta_c + \epsilon \geq m_c/M + 1/M +\epsilon/2 = (m_c+1)/M + \epsilon/2,
\end{eqnarray*}
where the middle inequality holds because $\theta_c \geq m_c/M$ and we chose $M$ such that $\epsilon/2 > 1/M$. In this case, college $c$ belongs in one of the sets on the second line of (\ref{eq:number-nontarget-schools-ineq}), namely the summation term for $m=m_c+1$.
 If (ii) holds, then
\begin{eqnarray*}
\rho(c) \leq \theta_c - \epsilon < m_c/M-\epsilon/2,
\end{eqnarray*}
in which case, school $c$ belongs to one of the sets on the first line of (\ref{eq:number-nontarget-schools-ineq}), namely the summation term for $m=m_c$. 

Therefore, each school $c$ belongs to one of the sets on the right-hand side of (\ref{eq:number-nontarget-schools-ineq}), and so inequality (\ref{eq:number-nontarget-schools-ineq}) holds. Now, by Propositions \ref{loser-schools-go-to-0} and \ref{winner-schools-go-to-0} and the weak law of large numbers, the right-hand side of (\ref{eq:number-nontarget-schools-ineq}) converges to $1/M$, which is strictly less than $\delta$, which shows equation (\ref{eq:convergence-non-target-schools}) and completes the proof of Theorem \ref{thm:correct-rankings-limit}.

\label{last-page-appendix}

\end{document}

%% file: figures/IRUS_desirability_intro.tex
\begin{figure}
\centering
\caption{\bf Graphical depiction of IRUS using a desirability graph}

\begin{subfigure}[t]{0.32\textwidth}
  \centering
  \begin{minipage}[t][4.5cm][t]{\linewidth}

    \begin{tikzpicture}[scale=0.5, baseline=(current bounding box.north)]

\tikzstyle{student1} = [circle,draw=black,fill=bg_comp2,thick, inner sep=0pt,minimum size=.75cm]
\tikzstyle{student2} = [circle,draw=black,fill=bg_comp1,thick, inner sep=0pt,minimum size=.75cm]
\tikzstyle{school}=[rectangle, fill=beach_green,
		inner sep=7pt,draw, minimum size = 2cm]
\tikzstyle{seat}=[rectangle, fill=beach_green,
		inner sep=3pt,draw, minimum size = .75cm]
\tikzstyle{normal} = [thick, line width = 3pt]
\tikzstyle{point}= [thick,color = bg_comp1, line width=3pt]

\Tball{s_blue}{blue}{2}
\Tball{m_red}{red}{3}

\foreach \x/ \y / \z in {1/2/10, 2/5/10,3/6/10,4/7/10, 5/9/10, 6/10/10, 7/1/8, 8/3/8, 9/8/8, 10/1/6, 11/3/6, 12/8/6, 13/2/4, 14/8/4, 15/3/2,16/5/2, 17/9/2}
	\node[s_blue](c\x) at (\y,\z){};

\tikzstyle{edge}= [->,line width = 1pt]
\foreach \x / \y in {10/7,15/13,15/1,15/11, 16/12, 16/14, 17/14, 17/12, 17/9, 17/5}
	\draw[edge](c\x)--(c\y);
\foreach \x / \y in {7/1, 7/2,13/11, 13/9, 14/1, 14/3, 14/12}
	\draw[edge](c\x)--(c\y);

\foreach \x / \y in { 11/8, 11/3,12/9}
	\draw[edge](c\x)--(c\y);
\foreach \x / \y in {8/1,9/4, 9/5}
	\draw[edge](c\x)--(c\y);

    \end{tikzpicture}

\vfill

  \end{minipage}
    \caption{Initial desirability graph}\label{subfig:initial-desire-graph}
\end{subfigure}
\hfill
\begin{subfigure}[t]{0.32\textwidth}
  \centering
  \begin{minipage}[t][4.5cm][t]{\linewidth}

    \begin{tikzpicture}[scale=0.5, baseline=(current bounding box.north)]

\tikzstyle{student1} = [circle,draw=black,fill=bg_comp2,thick, inner sep=0pt,minimum size=.75cm]
\tikzstyle{student2} = [circle,draw=black,fill=bg_comp1,thick, inner sep=0pt,minimum size=.75cm]
\tikzstyle{school}=[rectangle, fill=beach_green,
		inner sep=7pt,draw, minimum size = 2cm]
\tikzstyle{seat}=[rectangle, fill=beach_green,
		inner sep=3pt,draw, minimum size = .75cm]
\tikzstyle{normal} = [thick, line width = 3pt]
\tikzstyle{point}= [thick,color = bg_comp1, line width=3pt]

\Tball{s_blue}{blue}{2}
\Tball{m_red}{red}{3}
\tikzstyle{red_star}=[shape = star, star points = 7,%
		shading = ball,%
		very thin,
		]

\foreach \x/ \y / \z in {1/2/10, 2/5/10,3/6/10,4/7/10, 5/9/10, 6/10/10, 7/1/8, 8/3/8, 9/8/8, 10/1/6, 11/3/6, 12/8/6, 13/2/4, 14/8/4, 15/3/2,16/5/2, 17/9/2}
	\node[s_blue](c\x) at (\y,\z){};

\tikzstyle{edge}= [->,line width = 1pt]
\foreach \x / \y in {10/7,15/13,15/1,15/11, 16/12, 16/14, 17/14, 17/12, 17/9, 17/5}
	\draw[edge](c\x)--(c\y);
\foreach \x / \y in {7/1, 7/2,13/11, 13/9, 14/1, 14/3, 14/12}
	\draw[edge](c\x)--(c\y);

\foreach \x / \y in { 11/8, 11/3,12/9}
	\draw[edge](c\x)--(c\y);
\foreach \x / \y in {8/1,9/4, 9/5}
	\draw[edge](c\x)--(c\y);

\foreach \x/ \y / \z in {10/1/6, 15/3/2, 16/5/2, 17/9/2, 6/10/10}
	\node[star, star points=7, minimum height=.2cm, fill=red](c\x) at (\y,\z){};

    \end{tikzpicture}

\vfill

  \end{minipage}
      \caption{Last-ranked schools}\label{subfig:desire-graph-last-cycles}
\end{subfigure}
\hfill
\begin{subfigure}[t]{0.32\textwidth}
  \centering
  \begin{minipage}[t][4.5cm][t]{\linewidth}

    \begin{tikzpicture}[scale=0.5, baseline=(current bounding box.north)]

\tikzstyle{student1} = [circle,draw=black,fill=bg_comp2,thick, inner sep=0pt,minimum size=.75cm]
\tikzstyle{student2} = [circle,draw=black,fill=bg_comp1,thick, inner sep=0pt,minimum size=.75cm]
\tikzstyle{school}=[rectangle, fill=beach_green,
		inner sep=7pt,draw, minimum size = 2cm]
\tikzstyle{seat}=[rectangle, fill=beach_green,
		inner sep=3pt,draw, minimum size = .75cm]
\tikzstyle{normal} = [thick, line width = 3pt]
\tikzstyle{point}= [thick,color = bg_comp1, line width=3pt]

\Tball{s_blue}{blue}{2}
\Tball{m_red}{red}{3}

\foreach \x/ \y / \z in {1/2/10, 2/5/10, 3/6/10, 4/7/10, 5/9/10,  7/1/8, 8/3/8, 9/8/8,  11/3/6, 12/8/6, 13/2/4, 14/8/4 }
	\node[s_blue](c\x) at (\y,\z){};

\tikzstyle{edge}= [->,line width = 1pt]
\foreach \x / \y in {7/1, 7/2,13/11, 13/9, 14/1, 14/3, 14/12}
	\draw[edge](c\x)--(c\y);

\foreach \x / \y in { 11/8, 11/3,12/9}
	\draw[edge](c\x)--(c\y);
\foreach \x / \y in {8/1,9/4, 9/5}
	\draw[edge](c\x)--(c\y);

\foreach \x/ \y / \z in {13/2/4, 14/8/4, 7/1/8}
	\node[star, star points=7, minimum height=.2cm, fill=red](c\x) at (\y,\z){};
    \end{tikzpicture}

  \vfill
    
  \end{minipage}
  \caption{2nd-to-last-ranked schools}\label{subfig:desire-graph-second-to-last-schools}
\end{subfigure}
\label{fig:desire-graph-intro}
\end{figure}

%% file: figures/example_preference_rankings.tex
\begin{figure}[h]
\centering
 \caption{\bf Example of students' preference rankings \label{fig:quality-level-example}}
    \begin{tabular}{c|c|c|c}
    \toprule
        $i_1$ & $i_2$ & $i_3$ & $i_4$ \\
        \midrule
        $H_1^*$ & $H_2^*$ & $H_1$ & $H_2$ \\
        \fbox{$H_2$} & \fbox{$H_1$} & $H_2$ & $H_1$ \\
        $L_1$ & $L_1$ & $L_1^*$ & $L_2^*$ \\
        $L_2$ & $L_2$ & \fbox{$L_2$} & \fbox{$L_1$} \\
        \bottomrule
    \end{tabular}
    \begin{figurenotes}
        Each column indicates the preference ranking of a student. The boxes denote the initial matching $\mu$, and the stars are the shadow matching $\mu^*$.
       \end{figurenotes}
        \end{figure}

%% file: figures/illustration_justification_axiom.tex
\begin{figure}[h]
    \centering
        \caption{\bf Illustration of the justification axiom \label{tab:jusitified-example}}

    \begin{tabular}{c|c|c|c}
        \toprule
        $P_1$ & $P_2$ & $P_3$ & $P_4$ \\
        \midrule
        $\boxed{A}$ & $A$ & $A$ & $\boxed{D}$ \\
        $B$ & $\boxed{B}$ & $B$ & $A$ \\
        $C$ & $C$ & $\boxed{C}$ & $B$ \\
        $D$ & $D$ & $D$ & $C$ \\
        \bottomrule

    \end{tabular}
\end{figure}

%% file: figures/IRUS_desirability_text.tex
\begin{figure}[h]
\centering
\caption{\bf Graphical depiction of IRUS using a desirability graph}

\begin{subfigure}[t]{0.32\textwidth}
  \centering
  \begin{minipage}[t][4.5cm][t]{\linewidth}
\caption{Initial desirability graph}\label{subfig:initial-desire-graph}

    \begin{tikzpicture}[scale=0.5, baseline=(current bounding box.north)]

\tikzstyle{student1} = [circle,draw=black,fill=bg_comp2,thick, inner sep=0pt,minimum size=.75cm]
\tikzstyle{student2} = [circle,draw=black,fill=bg_comp1,thick, inner sep=0pt,minimum size=.75cm]
\tikzstyle{school}=[rectangle, fill=beach_green,
		inner sep=7pt,draw, minimum size = 2cm]
\tikzstyle{seat}=[rectangle, fill=beach_green,
		inner sep=3pt,draw, minimum size = .75cm]
\tikzstyle{normal} = [thick, line width = 3pt]
\tikzstyle{point}= [thick,color = bg_comp1, line width=3pt]

\Tball{s_blue}{blue}{2}
\Tball{m_red}{red}{3}

\foreach \x/ \y / \z in {1/2/10, 2/5/10,3/6/10,4/7/10, 5/9/10, 6/10/10, 7/1/8, 8/3/8, 9/8/8, 10/1/6, 11/3/6, 12/8/6, 13/2/4, 14/8/4, 15/3/2,16/5/2, 17/9/2}
	\node[s_blue](c\x) at (\y,\z){};

\tikzstyle{edge}= [->,line width = 1pt]
\foreach \x / \y in {10/7,15/13,15/1,15/11, 16/12, 16/14, 17/14, 17/12, 17/9, 17/5}
	\draw[edge](c\x)--(c\y);
\foreach \x / \y in {7/1, 7/2,13/11, 13/9, 14/1, 14/3, 14/12}
	\draw[edge](c\x)--(c\y);

\foreach \x / \y in { 11/8, 11/3,12/9}
	\draw[edge](c\x)--(c\y);
\foreach \x / \y in {8/1,9/4, 9/5}
	\draw[edge](c\x)--(c\y);

    \end{tikzpicture}

\vfill

  \end{minipage}
\end{subfigure}
\hfill
\begin{subfigure}[t]{0.32\textwidth}
  \centering
   \caption{Last-ranked schools}\label{subfig:desire-graph-last-cycles}
  \begin{minipage}[t][4.5cm][t]{\linewidth}

    \begin{tikzpicture}[scale=0.5, baseline=(current bounding box.north)]

\tikzstyle{student1} = [circle,draw=black,fill=bg_comp2,thick, inner sep=0pt,minimum size=.75cm]
\tikzstyle{student2} = [circle,draw=black,fill=bg_comp1,thick, inner sep=0pt,minimum size=.75cm]
\tikzstyle{school}=[rectangle, fill=beach_green,
		inner sep=7pt,draw, minimum size = 2cm]
\tikzstyle{seat}=[rectangle, fill=beach_green,
		inner sep=3pt,draw, minimum size = .75cm]
\tikzstyle{normal} = [thick, line width = 3pt]
\tikzstyle{point}= [thick,color = bg_comp1, line width=3pt]

\Tball{s_blue}{blue}{2}
\Tball{m_red}{red}{3}
\tikzstyle{red_star}=[shape = star, star points = 7,%
		shading = ball,%
		very thin,
		]

\foreach \x/ \y / \z in {1/2/10, 2/5/10,3/6/10,4/7/10, 5/9/10, 6/10/10, 7/1/8, 8/3/8, 9/8/8, 10/1/6, 11/3/6, 12/8/6, 13/2/4, 14/8/4, 15/3/2,16/5/2, 17/9/2}
	\node[s_blue](c\x) at (\y,\z){};

\tikzstyle{edge}= [->,line width = 1pt]
\foreach \x / \y in {10/7,15/13,15/1,15/11, 16/12, 16/14, 17/14, 17/12, 17/9, 17/5}
	\draw[edge](c\x)--(c\y);
\foreach \x / \y in {7/1, 7/2,13/11, 13/9, 14/1, 14/3, 14/12}
	\draw[edge](c\x)--(c\y);

\foreach \x / \y in { 11/8, 11/3,12/9}
	\draw[edge](c\x)--(c\y);
\foreach \x / \y in {8/1,9/4, 9/5}
	\draw[edge](c\x)--(c\y);

\foreach \x/ \y / \z in {10/1/6, 15/3/2, 16/5/2, 17/9/2, 6/10/10}
	\node[star, star points=7, minimum height=.2cm, fill=red](c\x) at (\y,\z){};

    \end{tikzpicture}

\vfill

  \end{minipage}
\end{subfigure}
\hfill
\begin{subfigure}[t]{0.32\textwidth}
  \centering
  \begin{minipage}[t][4.5cm][t]{\linewidth}
  \caption{2nd-to-last-ranked schools}\label{subfig:desire-graph-second-to-last-schools}

    \begin{tikzpicture}[scale=0.5, baseline=(current bounding box.north)]

\tikzstyle{student1} = [circle,draw=black,fill=bg_comp2,thick, inner sep=0pt,minimum size=.75cm]
\tikzstyle{student2} = [circle,draw=black,fill=bg_comp1,thick, inner sep=0pt,minimum size=.75cm]
\tikzstyle{school}=[rectangle, fill=beach_green,
		inner sep=7pt,draw, minimum size = 2cm]
\tikzstyle{seat}=[rectangle, fill=beach_green,
		inner sep=3pt,draw, minimum size = .75cm]
\tikzstyle{normal} = [thick, line width = 3pt]
\tikzstyle{point}= [thick,color = bg_comp1, line width=3pt]

\Tball{s_blue}{blue}{2}
\Tball{m_red}{red}{3}

\foreach \x/ \y / \z in {1/2/10, 2/5/10, 3/6/10, 4/7/10, 5/9/10,  7/1/8, 8/3/8, 9/8/8,  11/3/6, 12/8/6, 13/2/4, 14/8/4 }
	\node[s_blue](c\x) at (\y,\z){};

\tikzstyle{edge}= [->,line width = 1pt]
\foreach \x / \y in {7/1, 7/2,13/11, 13/9, 14/1, 14/3, 14/12}
	\draw[edge](c\x)--(c\y);

\foreach \x / \y in { 11/8, 11/3,12/9}
	\draw[edge](c\x)--(c\y);
\foreach \x / \y in {8/1,9/4, 9/5}
	\draw[edge](c\x)--(c\y);

\foreach \x/ \y / \z in {13/2/4, 14/8/4, 7/1/8}
	\node[star, star points=7, minimum height=.2cm, fill=red](c\x) at (\y,\z){};
    \end{tikzpicture}

  \vfill
    
  \end{minipage}
\end{subfigure}
\label{fig:desire-graph}
\end{figure}

%% file: figures/example_IRUS_algo.tex
\begin{figure}[h]
\centering
\caption{\bf Example of the IRUS algorithm\label{fig:Example-1}}
\makebox[\textwidth][c]{%
  \subcaptionbox{Step 1 (entire market)\label{fig:irus-step1}}[0.3\textwidth]{%
    \centering
    \[
    \begin{array}{c|c|c|c|c|c}
    \toprule
    P_1 & P_2 & P_3 & P_4 & P_5 & P_6 \\ \midrule
    A & C & B & A^* & B^* & \boxed{F^*} \\
    B & A & \boxed{C^*} & \boxed{B} & \boxed{A} & A \\
    D^* & E^* & A & C & C & B \\
    \boxed{E} & \boxed{D} & D & D & E & C \\
    C & B & E & E & D & D \\
    F & F & F & F & F & E\\
    \bottomrule
    \end{array}
  \]
  }
  \hspace{3em}
  \subcaptionbox{Step 2\label{fig:irus-step2}}[0.3\textwidth]{%
    \centering
    \[
    \begin{array}{c|c|c}
        \toprule
    P_3 & P_4 & P_5 \\ \midrule
    B & A^* & B^* \\
    \boxed{C^*} & \boxed{B} & \boxed{A} \\
    A & C & C\\
    \bottomrule
    \end{array}
    \]
    \vspace{3.8em} 
  }
  \hspace{-3em}
  \subcaptionbox{Step 3\label{fig:irus-step3}}[0.3\textwidth]{%
    \centering
    \[
    \begin{array}{c}
    \toprule
    P_5 \\ \midrule
    B^* \\
    \bottomrule
 \\
   \\
    \end{array}
    \]
    \vspace{4em} 
  }
}
\end{figure}

%% file: figures/random_graph.tex
\begin{figure}
\begin{center}
\caption{\bf Schematic of a random graph\label{fig:random-graph}}
\begin{tikzpicture}[scale=0.7]

\draw (0,0) -- (16,0);
\node at (0,0) [below] {};
\node at (16,0) [below] {};

\foreach \x/\label in {0.08/, 0.16/, 0.35/, 0.4/, 0.44/, 0.6/, 0.75/, 0.95/}
    \node (\label) at (\x*16,0) [circle, fill=black, inner sep=2pt] {};

\foreach \x/\label in {0.25/\Large$a$, 0.50/\Large$b$} {
    \draw[thick] (\x*16,0.2) -- (\x*16,-0.2);
    \node at (\x*16,0.3) [above] {\label};
}

\foreach \x/\label in {0.08/, 0.16/, 0.35/, 0.4/, 0.44/, 0.6/, 0.75/, 0.95/}
    \node at (\x*16,-4) [circle, fill=black, inner sep=2pt] {};

\foreach \x in {0.08, 0.16}{
	\foreach \y in {0.6,0.75, 0.95}{
	\draw (\x*16,-4) -- (\y*16,0);
	
	}
	
}

\foreach \x in {0.16, 0.4}
	\draw (.08*16,-4) -- (\x*16,0);
	
\foreach \x in {0.08, 0.35, 0.6}
	\draw (0.4*16,-4) -- (\x*16,0);
	
\foreach \x in {0.16, 0.35, 0.4,0.95}
	\draw(0.6*16,-4) -- (\x*16,0); 

\foreach \x in {0.44, 0.75}
	\draw(0.95*16,-4) -- (\x*16,0); 

\foreach \x/\label in {0.08/, 0.16/, 0.35/, 0.4/, 0.44/, 0.6/, 0.75/, 0.95/}
	\draw[dashed, dash pattern=on 3pt off 2pt] (\x*16,0) -- (\x*16,-4);


\draw[red, thick] (1.9,-4) ellipse (1cm and 0.3cm);

\draw[red, thick] (12.4,0) ellipse (3.2cm and 0.5cm);

\node [font=\Huge] at (2, 0.8) {$L$};

\node [font=\Huge, color=red] at (12.2, 1) {$H$};
\node [font=\Large] at (16.5,0) {$\theta_c$};
\node [font=\Huge, color=red] at (0.3,-4) {$U$};
\node [font=\Large] at (16.5,-4) {$I$};
\end{tikzpicture}
\begin{figurenotes}
The dashed lines represent the shadow matching $\mus$, while the solid lines are the bipartite graph $G^p$, where student $i$ and college $c$ are connected if $\eta_{i,c}< p = 1-\Delta$. The set $U \cup H$ is a biclique, and the size of the maximum balanced biclique is $\min\{|U|,|H|\}=2$.
\end{figurenotes}
\end{center}
\end{figure}

%% file: figures/scatter_average_bias.tex
\begin{figure}[t!!]
			\caption{\bf Simulation Results: Desriable Ranking vs. True Rankings\label{fig:scatter_rho_true}}
	  \hspace{-0.4in} \includegraphics[scale=0.6]{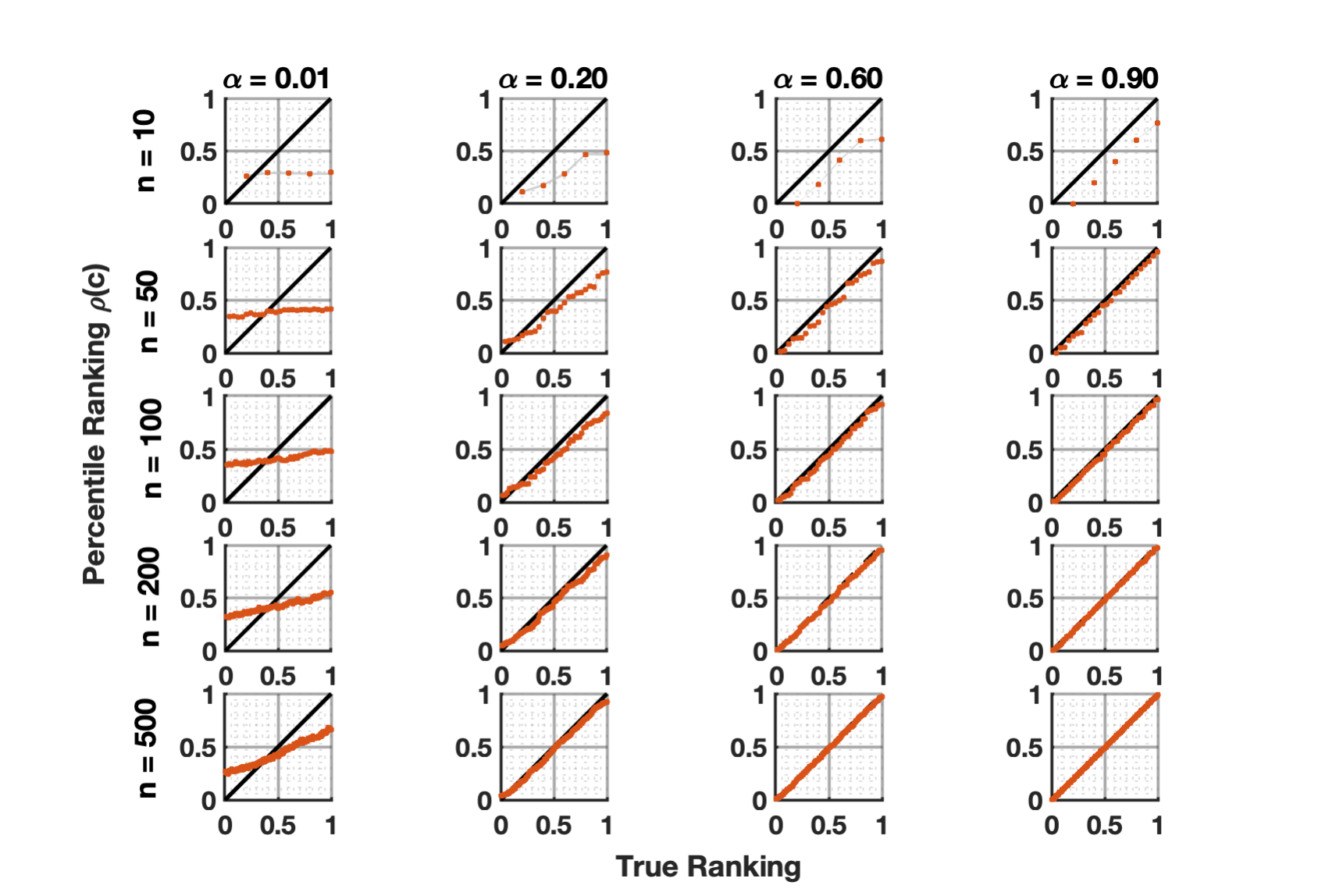}	
				\begin{figurenotes}
				The figure displays the relationship between true college rankings (x-axis) and the percentile ($\rho$) ranking (y-axis) across different market sizes ($n$) and preference concentration parameters ($\alpha$). Each point represents the average $\rho$-ranking across 1,000 simulations for a single college, using the randomly selected 93rd shadow matching $\mu^*$. The solid diagonal line indicates perfect ranking accuracy ($\rho = $ true rank). 		\end{figurenotes}
	\end{figure}

%% file: tables/simulation_bias.tex
\begin{sidewaystable}[htbp]
\caption{\bf Simulation Results: Performance of Desirable Rankings\label{tab:bias}}\vspace{0.2in}
\scalebox{0.8}{
\begin{tabular}{l@{\hspace{8pt}}ccccc@{\hspace{12pt}}c@{\hspace{12pt}}ccccc}
\toprule
          & \multicolumn{5}{l}{$\alpha=0.01$}           &  & \multicolumn{5}{l}{$\alpha=0.20$}  \\
          \cline{2-6} \cline{8-12}          \\
          & $n=10$ & $50$ & $100$ & $200$ & $500$ &  & $10$ & $50$ & $100$ & $200$ & $500$ \\
\midrule
Bias      & 0.446    & 0.271    & 0.227    & 0.183    & 0.133    &  & 0.272    & 0.098    & 0.051    & 0.040    & 0.054    \\
          & [0.444,0.452]    & [0.266,0.275]    & [0.225,0.229]    & [0.182,0.186]    & [0.132,0.134]    &  & [0.266,0.277]    & [0.096,0.099]    & [0.050,0.052]    & [0.040,0.041]    & [0.054,0.054]    \\
Var      & 0.079    & 0.081    & 0.079    & 0.077    & 0.065    &  & 0.055    & 0.036    & 0.023    & 0.013    & 0.005    \\
          & [0.076,0.080]    & [0.080,0.081]    & [0.079,0.080]    & [0.076,0.077]    & [0.064,0.065]    &  & [0.054,0.057]    & [0.035,0.036]    & [0.023,0.023]    & [0.013,0.013]    & [0.005,0.006]    \\
MSE      & 0.313    & 0.175    & 0.151    & 0.125    & 0.088    &  & 0.166    & 0.047    & 0.026    & 0.015    & 0.009    \\
          & [0.309,0.318]    & [0.170,0.176]    & [0.150,0.153]    & [0.124,0.126]    & [0.087,0.089]    &  & [0.164,0.172]    & [0.047,0.049]    & [0.026,0.026]    & [0.015,0.015]    & [0.009,0.009]    \\
\midrule
          & \multicolumn{5}{l}{$\alpha=0.60$}           &  & \multicolumn{5}{l}{$\alpha=0.90$}  \\\cline{2-6} \cline{8-12}          \\
          & $n=10$ & $50$ & $100$ & $200$ & $500$ &  & $10$ & $50$ & $100$ & $200$ & $500$ \\
\midrule
Bias      & 0.312    & 0.086    & 0.029    & 0.062    & 0.039    &  & 0.252    & 0.036    & 0.048    & 0.017    & 0.023    \\
          & [0.310,0.314]    & [0.085,0.087]    & [0.028,0.029]    & [0.062,0.062]    & [0.039,0.039]    &  & [0.251,0.253]    & [0.035,0.036]    & [0.048,0.048]    & [0.017,0.017]    & [0.023,0.023]    \\
Var      & 0.012    & 0.006    & 0.003    & 0.002    & 0.001    &  & 0.001    & 0.000    & 0.000    & 0.000    & 0.000    \\
          & [0.011,0.012]    & [0.006,0.006]    & [0.003,0.003]    & [0.002,0.002]    & [0.001,0.001]    &  & [0.001,0.001]    & [0.000,0.001]    & [0.000,0.001]    & [0.000,0.000]    & [0.000,0.000]    \\
MSE      & 0.116    & 0.013    & 0.005    & 0.007    & 0.003    &  & 0.073    & 0.002    & 0.004    & 0.001    & 0.001    \\
          & [0.115,0.117]    & [0.013,0.014]    & [0.004,0.005]    & [0.007,0.007]    & [0.003,0.003]    &  & [0.072,0.073]    & [0.002,0.002]    & [0.004,0.004]    & [0.001,0.001]    & [0.001,0.001]    \\
\bottomrule
\end{tabular}}
\begin{tablenotes}[flushleft]
\item \footnotesize \emph{Note:} This table reports the performance of desirable rankings $\rho$ under different sample sizes and preference weights. 
The top numbers in each cell represents the bias, variance and the mean squared error (MSE) calculated using one randomly selected shadow matching (93rd out of 100) across 1,000 Monte Carlo simulations. The numbers in the brackets shown below each entry is the range across all 100 shadow assignments. 
\end{tablenotes}
\end{sidewaystable}

%% file: figures/simulation_uniform_convergence.tex
\begin{figure}[ht!!]
			\caption{\bf Simulation Results: Uniform Convergence of Desirable Rankings\label{fig:uniform_convergence}}
\hspace{-0.85in}	   \includegraphics[scale=0.45]{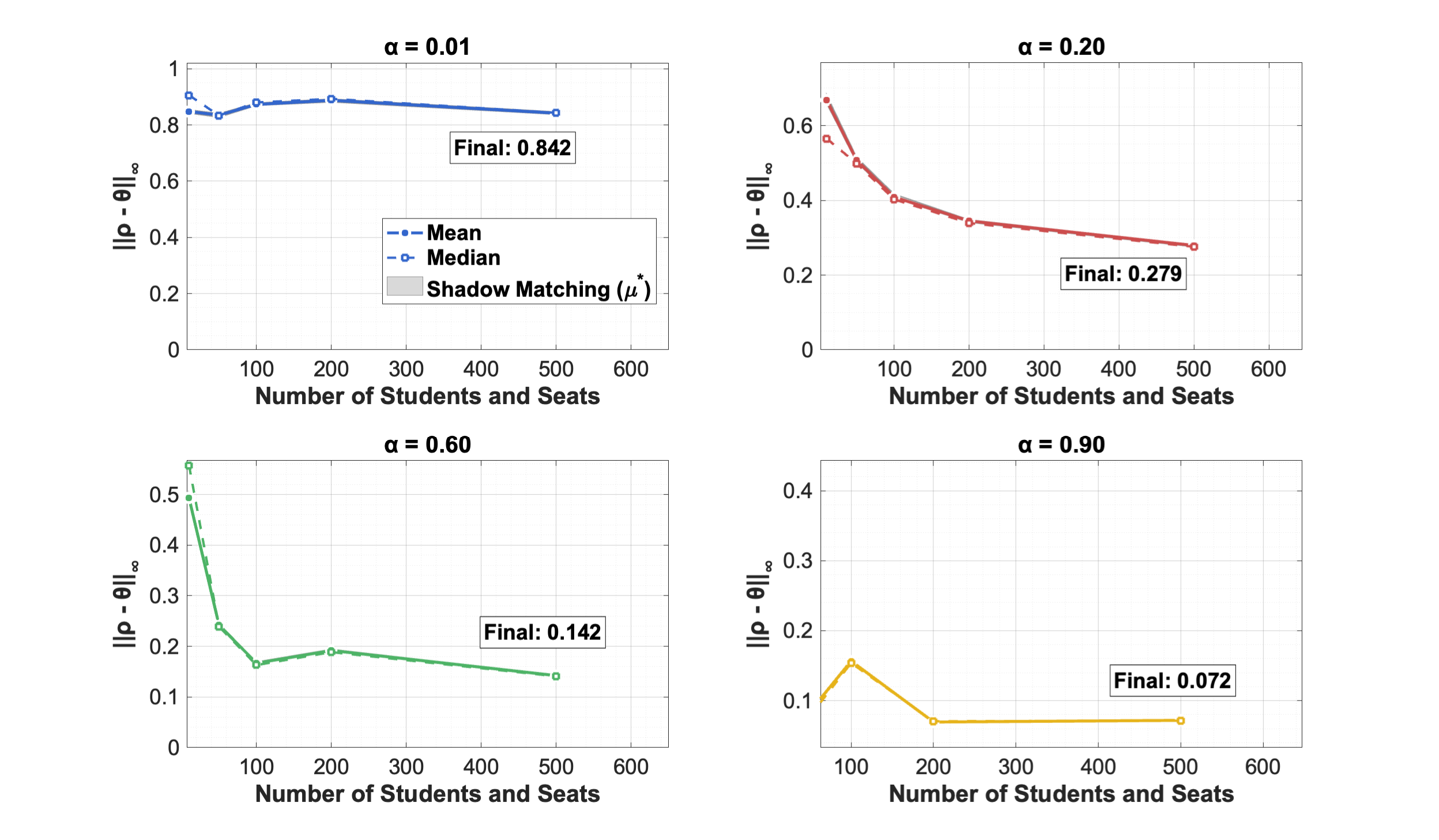}	
				\begin{figurenotes}
		The figure shows uniform convergence of desirable rankings $\|\rho - \theta\|_{\infty}$ across market sizes ($n$) for different preference weights ($\alpha$). Market size is defined as the number of students (and the number of seats). Solid and dashed lines represent mean and median convergence using one out of a hundred randomly selected $\mu^*$. Gray bands show the point-wise minimum and maximum across all 100 Pareto improvement methods. (These bands are difficult to see in the figure because they are very tight, which is a demonstration of the robustness to the choice of $\mu^*$.)
		\end{figurenotes}
	\end{figure}

%% file: tables/data_summary_stat.tex
 \begin{sidewaystable}[htbp]
     \centering\vspace{0.3in}
     \caption{\bf Summary Statistics\label{tab:summary_statistics}}
\scalebox{0.9}{     
     \begin{tabular}{lcccc}
    \toprule
      & 2006 & 2007 & 2008 & 2009 \\
    \midrule
     \multicolumn{5}{l}{\textbf{Panel A: Students}} \\
     PSU Score & 6,797 (866) & 6,749 (888) & 6,850 (829) & 6,768 (825) \\
     Civil Status & Married (99.2\%) & Married (99.2\%) & Married (99.6\%) & Married (99.6\%) \\
     Student Region & Santiago Metro (37.3\%) & Santiago Metro (38.8\%) & Santiago Metro (41.1\%) & Santiago Metro (39.4\%) \\
     School Region & Santiago Metro (35.3\%) & Santiago Metro (36.8\%) & Santiago Metro (39.6\%) & Santiago Metro (38.1\%) \\
     Enrollment Rate & 44.5\% & 39.7\% & 42.4\% & 37.6\% \\
     Family Income & CLP\$278-834K (34.1\%) & CLP\$278-834K (34.9\%) & CLP\$1,080K+ (32.2\%) & CLP\$1,584K+ (21.2\%) \\
     Father Education & University Deg. (45.2\%) & University Deg. (47.5\%) & University Deg. (45.8\%) & University Deg. (44.2\%) \\
     Female & 47.2\% & 49.7\% & 51.6\% & 53.1\% \\
     Head of Family & Applicant's father (67.7\%) & Applicant's father (66.2\%) & Applicant's father (68.2\%) & Applicant's father (64.3\%) \\
     Health Coverage & Private (57.5\%) & Private (56.2\%) & Private (56.7\%) & Private (57.6\%) \\
     Mother Education & University Deg. (38.2\%) & University Deg. (41.3\%) & University Deg. (40.2\%) & University Deg. (37.7\%) \\
     Municipal School & 18.0\% & 19.0\% & 16.3\% & 15.8\% \\
     Num Applications & 3.91 (2.35) & 3.79 (2.41) & 3.43 (2.19) & 3.54 (2.16) \\
     Paid Work & No (90.1\%) & No (92.2\%) & No (93.5\%) & No (92.3\%) \\
     Private School & 47.0\% & 42.2\% & 44.4\% & 44.0\% \\
     PEM Score (Std.) & 708.1 (76.5) & 696.5 (96.3) & 701.3 (91.2) & 693.8 (87.7) \\
     Which Ranked College & 2.30 (1.56) & 1.92 (1.35) & 1.69 (1.11) & 1.96 (1.29) \\
     \midrule
     \multicolumn{5}{l}{\textbf{Panel B: Colleges}} \\
     Average Cutoff Score & 71,563 (3139) & 71,563 (3139) & 71,631 (3205) & 71,610 (3066) \\
     First Cutoff Score & 71,717 (3092) & 71717 (3092) & 71772 (3162) & 71,768 (3030) \\
     Last Cutoff Score & 71,409 (3191) & 71409 (3191) & 71490 (3252) & 71453 (3108) \\
     Central Location & 0.29 (0.46) & 0.29 (0.46) & 0.30 (0.47) & 0.32 (0.48) \\
     North Location & 0.10 (0.30) & 0.10 (0.30) & 0.05 (0.22) & 0.05 (0.21) \\
     South Location & 0.62 (0.50) & 0.62 (0.50) & 0.65 (0.49) & 0.64 (0.49) \\
     Selectivity &&&&\\
     -Cutoff-based & 0.01 (1.04) & 0.01 (1.04) & 0.04 (1.06) & 0.02 (1.02) \\
     -Preference-based & -0.04 (1.00) & -0.04 (1.00) & -0.00 (1.02) & 0.03 (1.01) \\
     -Student Quality & 0.01 (1.03) & 0.01 (1.03) & 0.03 (1.06) & 0.01 (1.02) \\
     \bottomrule
     \end{tabular}}
     \begin{tablenotes}
     \item \footnotesize{\textbf{Notes:} 
     Summary statistics for Chilean medical school data from 2006 to 2009. Panel A reports student characteristics, and Panel B reports college characteristics. For continuous variables, means are shown with standard deviations in parentheses. For binary variables, percentages are reported. For categorical variables, the most common category is shown with its percentage in parentheses. Family income is reported in thousands of Chilean pesos with varying bracket definitions across years. The three selectivity measures (see the text for definition) are standardized to have a mean of zero and unit variance, with higher values indicating more selective colleges. Sample sizes are N=1,302 (2006), N=1,321 (2007), N=1,168 (2008), and N=1,620 (2009).}
     \end{tablenotes}
     \end{sidewaystable}

%% file: tables/data_medical_school_matches.tex
\begin{table}[t!!]
\caption{\bf Medical School Match Outcomes by Year\label{tab:match_outcomes}}
\hspace{-0.2in}\scalebox{0.9}{
\begin{tabular}{lcccccccc}
\toprule
 & \multicolumn{2}{c}{2006} & \multicolumn{2}{c}{2007} & \multicolumn{2}{c}{2008} & \multicolumn{2}{c}{2009} \\
Outcome & Count & Percent & Count & Percent & Count & Percent & Count & Percent \\
\midrule
1st choice & 236 & 41.04\% & 282 & 54.23\% & 299 & 60.53\% & 297 & 49.17\% \\
2nd choice & 146 & 25.39\% & 114 & 21.92\% & 109 & 22.06\% & 159 & 26.32\% \\
3rd choice & 83 & 14.43\% & 66 & 12.69\% & 55 & 11.13\% & 80 & 13.25\% \\
4th choice & 50 & 8.70\% & 25 & 4.81\% & 9 & 1.82\% & 31 & 5.13\% \\
5th choice & 34 & 5.91\% & 17 & 3.27\% & 16 & 3.24\% & 22 & 3.64\% \\
6th choice & 9 & 1.57\% & 8 & 1.54\% & 3 & 0.61\% & 10 & 1.66\% \\
7th choice & 12 & 2.09\% & 6 & 1.15\% & 3 & 0.61\% & 3 & 0.50\% \\
8th choice & 5 & 0.87\% & 2 & 0.38\% & 0 & 0.00\% & 2 & 0.33\% \\
\midrule
Total Admitted & 575 & 44.16\% & 520 & 39.36\% & 494 & 42.29\% & 604 & 37.28\% \\
No admission & 727 & 55.84\% & 801 & 60.64\% & 674 & 57.71\% & 1016 & 62.72\% \\
\midrule
Total students & 1302 &  & 1321 & & 1168 &  & 1620 &  \\
\midrule
Admitted, not enrolled & 660 &  & 709 &  & 606 &  & 915 & \\
\bottomrule
\end{tabular}}
\begin{tablenotes}[flushleft]
\item \footnotesize \emph{Note:} 
Match outcomes for students applying exclusively to medical programs in Chile's centralized admission system, 2006-2009. Percentages for choice outcomes (1st-8th choice) are calculated as a share of total admitted and enrolled students. ``No admission'' shows students who received no offers from any ranked medical program. Total students include all applicants to medical programs in each year.
``Admitted, not enrolled'' indicates students who received at least one admission offer but chose not to enroll in any medical program. \end{tablenotes}
\end{table}

%% file: tables/desirable_rankings.tex
\begin{table}[t!!]
\caption{\bf Desirable Rankings \label{tab:desirable_rankings}}
\hspace{-0.6in}\begin{threeparttable}
\begin{tabular}{lllllcccc}
\toprule
\textbf{ID} & \textbf{University} & \textbf{Field} & \textbf{Location} & \multicolumn{4}{c}{\textbf{Rankings}} \\
 & & & & \textbf{2006} & \textbf{2007} & \textbf{2008} & \textbf{2009} \\
\midrule
1183 & University of Chile & Medicine & Santiago & 2 & 2 & 2 & 2 \\
1195 & University of Chile & Dentistry & Santiago & 5 & 7 & 6 & 7 \\
1256 & Catholic University & Dentistry & Santiago & - & - & - & 9 \\
1258 & Catholic University & Medicine & Santiago & 1 & 1 & 1 & 1 \\
1386 & University of Concepci\'on & Medicine & Concepci\'on & 5 & 3 & 3 & 3 \\
1391 & University of Concepci\'on & Dentistry & Concepci\'on & 12 & 13 & 12 & 12 \\
1691 & University of Santiago de Chile & Medicine & Metro Region & 4 & 3 & 3 & 4 \\
1713 & Austral University of Chile & Medicine & Valdivia & 8 & 6 & 7 & 6 \\
1731 & Austral University of Chile & Medicine & Osorno & 9 & 12 & 10 & 9 \\
1789 & Austral University of Chile & Dentistry & Valdivia & 18 & 19 & 18 & 20 \\
1887 & North Catholic University & Medicine & Coquimbo & 12 & 13 & 16 & 12 \\
1939 & University of Valpara\'iso & Medicine & San Felipe & 11 & 10 & 7 & 12 \\
1940 & University of Valpara\'iso & Medicine & Valpara\'iso & 3 & 3 & 5 & 5 \\
1950 & University of Valpara\'iso & Dentistry & Valpara\'iso & 14 & 15 & 14 & 15 \\
2447 & University of Antofagasta & Medicine & Coloso & 14 & 15 & - & - \\
2448 & University of Antofagasta & Dentistry & Coloso  & 19 & 20 & 19 & 20 \\
3026 & University of La Frontera & Medicine & Temuco & 7 & 7 & 7 & 7 \\
3033 & University of La Frontera & Dentistry & Temuco & 16 & 17 & 16 & 19 \\
3420 & University of Talca & Medicine & Talca & - & - & - & 17 \\
3423 & University of Talca & Dentistry & Talca & 16 & 17 & 14 & 18 \\
3505 & Del Maule Catholic University & Medicine & Talca & 10 & 10 & 12 & 15 \\
3620 & Catholic University of the & Medicine & Concepci\'on & - & 7 & 11 & 9 \\
 & Most Holy Conception & &&&&& \\
\bottomrule
\end{tabular}
\begin{tablenotes}
\item \footnotesize \emph{Note:} This table presents desirability rankings obtained using the IRUS algorithm, applied separately for each year from 2006 to 2009. Programs are ordered by their official ID number. Lower numbers indicate higher desirability. Missing entries (-) indicate that the program was not available in the data for that particular year.
\end{tablenotes}
\end{threeparttable}
\end{table}

%% file: tables/revealed_preference_rankings.tex
\begin{table}[t!!]
\caption{\bf Revealed Preference Rankings\label{tab:revealed_preference}}
\hspace{-0.62in}\begin{threeparttable}
\begin{tabular}{llllcccc}
\toprule
\textbf{ID} & \textbf{University} & \textbf{Field} & \textbf{Location} & \multicolumn{4}{c}{\textbf{Rankings}} \\
 & & & & \textbf{2006} & \textbf{2007} & \textbf{2008} & \textbf{2009} \\
\midrule
1183 & University of Chile & Medicine & Santiago & 7 & 6 & 5 & 4 \\
1195 & University of Chile & Dentistry & Santiago & 3 & 20 & 14 & 7 \\
1256 & Catholic University & Dentistry & Santiago & - & - & - & 19 \\
1258 & Catholic University & Medicine & Santiago & 20 & 4 & 16 & 5 \\
1386 & University of Concepci\'on & Medicine & Concepci\'on & 11 & 12 & 17 & 9 \\
1391 & University of Concepci\'on & Dentistry & Concepci\'on & 21 & 19 & 20 & 16 \\
1691 & University of Santiago de Chile & Medicine & Metro Region & 13 & 18 & 8 & 21 \\
1713 & Austral University of Chile & Medicine & Valdivia & 4 & 21 & 4 & 11 \\
1731 & Austral University of Chile & Medicine & Osorno & 17 & 15 & 19 & 20 \\
1789 & Austral University of Chile & Dentistry & Valdivia & 19 & 7 & 3 & 8 \\
1887 & North Catholic University & Medicine & Coquimbo & 10 & 2 & 15 & 10 \\
1939 & University of Valpara\'iso & Medicine & San Felipe  & 5 & 1 & 12 & 22 \\
1940 & University of Valpara\'iso & Medicine & Valpara\'iso & 14 & 11 & 11 & 17 \\
1950 & University of Valpara\'iso & Dentistry & Valpara\'iso & 9 & 16 & 7 & 15 \\
2447 & University of Antofagasta & Medicine & Coloso  & 8 & 8 & - & - \\
2448 & University of Antofagasta & Dentistry & Coloso  & 6 & 14 & 18 & 6 \\
3026 & University of La Frontera & Medicine & Temuco & 1 & 10 & 1 & 3 \\
3033 & University of La Frontera & Dentistry & Temuco & 18 & 17 & 10 & 13 \\
3420 & University of Talca & Medicine & Talca & - & - & - & 2 \\
3423 & University of Talca & Dentistry & Talca & 2 & 3 & 9 & 14 \\
3505 & Del Maule Catholic University & Medicine & Talca & 16 & 5 & 2 & 1 \\
3620 & Catholic University of the & Medicine & Concepci\'on & 12 & 13 & 13 & 12 \\
 & Most Holy Conception &&&&&& \\
\bottomrule
\end{tabular}
\begin{tablenotes}
\item \footnotesize \emph{Note:} This table presents revealed preference rankings estimated using college fixed effects from a discrete choice model with student-specific consideration sets. Year-by-year rankings are estimated separately for each year from 2006 to 2009. Missing entry (-) indicates the program was not in our sample that year.
\end{tablenotes}
\end{threeparttable}
\end{table}

%% file: tables/borda_rankings.tex
\begin{table}[t!!]
\caption{\bf Borda Rankings \label{tab:borda_chile}}
\hspace{-0.6in}\begin{threeparttable}
\begin{tabular}{lllllcccc}
\toprule
\textbf{ID} & \textbf{University} & \textbf{Field} & \textbf{Location} & \multicolumn{4}{c}{\textbf{Rankings}} \\
 & & & & \textbf{2006} & \textbf{2007} & \textbf{2008} & \textbf{2009} \\
\midrule
1183 & University of Chile & Medicine & Santiago & 1 & 1 & 1 & 1 \\
1195 & University of Chile & Dentistry & Santiago & 11 & 11 & 9 & 8 \\
1256 & Catholic University & Dentistry & Santiago & - & - & - & 6 \\
1258 & Catholic University & Medicine & Santiago & 3 & 5 & 3 & 3 \\
1386 & University of Concepci\'on & Medicine & Concepci\'on & 2 & 2 & 2 & 2 \\
1391 & University of Concepci\'on & Dentistry & Concepci\'on & 13 & 12 & 11 & 13 \\
1691 & University of Santiago de Chile & Medicine & Metro Region & 4 & 4 & 4 & 4 \\
1713 & Austral University of Chile & Medicine & Valdivia & 6 & 6 & 6 & 7 \\
1731 & Austral University of Chile & Medicine & Osorno & 15 & 14 & 14 & 17 \\
1789 & Austral University of Chile & Dentistry & Valdivia & 19 & 18 & 17 & 19 \\
1887 & North Catholic University & Medicine & Coquimbo & 18 & 17 & 18 & 20 \\
1939 & University of Valpara\'iso & Medicine & San Felipe & 10 & 10 & 13 & 14 \\
1940 & University of Valpara\'iso & Medicine & Valpara\'iso & 5 & 3 & 5 & 5 \\
1950 & University of Valpara\'iso & Dentistry & Valpara\'iso & 14 & 13 & 12 & 15 \\
2447 & University of Antofagasta & Medicine & Coloso & 12 & 15 & - & - \\
2448 & University of Antofagasta & Dentistry & Coloso & 20 & 20 & 19 & 21 \\
3026 & University of La Frontera & Medicine & Temuco & 9 & 7 & 7 & 9 \\
3033 & University of La Frontera & Dentistry & Temuco & 16 & 19 & 16 & 18 \\
3420 & University of Talca & Medicine & Talca & - & - & - & 11 \\
3423 & University of Talca & Dentistry & Talca & 17 & 16 & 15 & 16 \\
3505 & Del Maule Catholic University & Medicine & Talca & 8 & 8 & 10 & 12 \\
3620 & Catholic University of the & Medicine & Concepci\'on & 7 & 9 & 8 & 10 \\
 & Most Holy Conception & &&&&& \\
\bottomrule
\end{tabular}
\begin{tablenotes}
\item \footnotesize \emph{Note:} This table presents Borda Count rankings calculated from student preference lists, applied separately for each year from 2006 to 2009. Each student was awarded points equal to their number of ranked programs ($m$ points for first choice, $(m-1)$ for second choice, and so on). Programs are ordered by program ID number. Missing entries (-) indicate that the program was not available in the data for that particular year.
\end{tablenotes}

\end{threeparttable}
\end{table}